\documentclass[journal]{IEEEtran}
\usepackage{amsmath}
\usepackage{amsfonts}
\usepackage{amssymb}
\usepackage{graphicx}
\usepackage[draft]{fixme}
\usepackage[pdfborder={0 0 0}]{hyperref}
\usepackage[english]{babel}
\usepackage{rotating}
\usepackage{listings}
\usepackage{xcolor}
\usepackage{enumerate}
\usepackage{textcomp}
\usepackage{booktabs}
\usepackage{wrapfig}
\usepackage{paralist}
\usepackage{tgheros}
\usepackage[labelformat=simple]{subcaption}
\usepackage[utf8]{inputenc}
\usepackage[inline]{enumitem}
\usepackage{acronym}
\usepackage{paralist}
\usepackage[sort,compress]{cite}
\usepackage{acronym}
\usepackage{csquotes}
\usepackage{backnaur}

\newcommand{\tint}{\mathtt{Int}}
\newcommand{\tintset}{\mathtt{OrderedSet\{Int\}}}
\newcommand{\tbool}{\mathtt{Bool}}
\newcommand{\treal}{\mathtt{Real}}
\newcommand{\trealset}{\mathtt{OrderedSet\{Real\}}}
\newcommand{\tterms}[2][\Delta]{\mathtt{TERM}(O\cup{#1})_{#2}}
\newcommand{\ttermssimple}[1][s]{\mathtt{TERM}_{#1}}

\newcommand{\sant}{\mbox{SAN-T}}

\newcommand{\sanscale}{.65}

\hypersetup{
    colorlinks = true,
    citecolor = blue,
    linkcolor = purple
}

\fxsetup {
    layout={inline}, 
    theme=color
}

\allowdisplaybreaks

\acrodef{AN}{Activity Network}
\acrodef{ADVANCE}{Addressing Verification and Validation Challenges in
Cyber-Physical Systems}
\acrodef{BEDCS}{Brazilian Environmental Data Collection System}
\acrodef{BVR}{Better Variability Resolution}
\acrodef{CCSL}{Coding Conventions Specification Language}
\acrodef{CD}{Continuous Delivery}
\acrodef{CI}{Continuous Integration}
\acrodef{CICD}[CI/CD]{\acl{CI}/\acl{CD}}
\acrodef{CPN}{Colored Petri Net}
\acrodef{CPS}{Cyber-Physical System}
\acrodef{CPSoS}{Cyber-Physical System-of-Systems}
\acrodef{CS}{Constituent System}
\acrodef{CVL}{Common Variability Language}
\acrodef{DARep}{Dependency-Aware Replication}
\acrodef{DCP}{Data Collection Platform}
\acrodef{DSL}{Domain-Specific Language}
\acrodef{EMF}{Eclipse Modeling Framework}
\acrodef{FMEA}{Failure Modes and Effects Analyis}
\acrodef{FT}{Fault Tree}
\acrodef{fUML}{Foundational \acs{UML}}
\acrodef{MBSE}{Model-Based Systems Engineering}
\acrodef{MDE}{Model-Driven Engineering}
\acrodef{O+M}[O\&M]{Operation \& Maintenance}
\acrodef{OMG}{Object Management Group}
\acrodef{PCD}{Plataforma de Coleta de Dados}
\acrodef{PN}{Petri Net}
\acrodef{RBD}{Reliability Block Diagram}
\acrodef{SAN}{Stochastic Activity Network}
\acrodef{SAN-T}{Stochastic Activity Network Template}
\acrodef{SDN}{Software Defined Network}
\acrodef{SBCDA}{Sistema Brasileiro de Coleta de Dados Ambientais}
\acrodef{SoS}{System-of-Systems}
\acrodef{SPL}{Software Product Line}
\acrodef{SPN}{Stochastic Petri Net}
\acrodef{SRN}{Stochastic Reward Net}
\acrodef{DSPL}{Dynamic Software Product Line}
\acrodef{TMDL}{Template Models Description Language}
\acrodef{UML}{Unified Modeling Language}
\acrodef{VV}[V\&V]{Verification \& Validation}

\begin{document}

\noindent
This work has been submitted to the IEEE for possible publication. Copyright may be transferred without notice, after which this version may no longer be accessible.

\title{Stochastic Activity Networks Templates: Supporting Variability in Performability Models}

\author{Leonardo Montecchi, Paolo Lollini, and Andrea Bondavalli\vspace{-5mm}
\thanks{L.~Montecchi is with the University of Campinas, Brazil; \texttt{leonardo@ic.unicamp.br}. 
	P.~Lollini and A.~Bondavalli are with the University of Firenze, Italy; \texttt{\{lollini,bondavalli\}@unifi.it}.}%
\thanks{This work has received funding from the European Union's Horizon
2020 research and innovation program under the Marie Sklodowska-Curie
grant agreement No 823788 ``ADVANCE''. This work has received funding from the
São Paulo Research Foundation (FAPESP) with grant \mbox{\#2019/02144-6}.}}

\maketitle

\begin{abstract}
Model-based evaluation is extensively used to estimate performance and
reliability of dependable systems. Traditionally, those systems were small and
self-contained, and the main challenge for model-based evaluation has been the
efficiency of the solution process. Recently, the problem of specifying and
maintaining complex models has increasingly gained attention, as modern systems
are characterized by many components and complex interactions. Components share
similarities, but also exhibit variations in their behavior due to different
configurations or roles in the system. From the modeling perspective, variations
lead to replicating and altering a small set of base models multiple times.
Variability is taken into account only informally, by defining a sample model
and explaining its possible variations. In this paper we address the problem of
including variability in performability models, focusing on Stochastic Activity
Networks (SANs). We introduce the formal definition of Stochastic Activity
Networks Templates (SAN-T), a formalism based on SANs with the addition of
variability aspects. Differently from other approaches, parameters can also
affect the structure of the model, like the number of cases of activities. We
apply the \mbox{SAN-T} formalism to the modeling of the backbone network of an
environmental monitoring infrastructure. In particular, we show how existing SAN
models from the literature can be generalized using the newly introduced
formalism.
\end{abstract}





\section{Introduction}

Formal methods have been extensively used to estimate performance and
reliability metrics of computer systems.
They are especially useful for assessing non-functional
properties of critical systems, for which experimental approaches are not always
applicable.
In fact, \emph{model-based evaluation} \cite{Nicol04} has the advantage of not
exercising the real system, which may be dangerous, costly, or
not feasible.

Traditionally, critical systems have been mostly isolated and
monolithic, and the main challenge for model-based evaluation has always been
the solution process, in term of efficiency of state-space generation
and accuracy of results. More recently, the problem of
\emph{specifying} and \emph{maintaining} complex models in convenient ways has
increasingly gained attention.
Modularization is an established approach in reducing the complexity in the
specification of analysis models. However, one of the rising challenges consists
in handling variability \cite{Coplien98,vanGurp01} across system components.

\ac{CPSoS} \cite{AMADEOSBook} are characterized by a large number of components
and complex interactions between them.
Many of these elements share similarities, but at the same time they have a
slight different behavior due to their individual configuration or role in the system.
When modeling the entire system, these variations  lead to
replicating and altering a small set of base models multiple times.
Furthermore, due to dynamicity and evolution \cite{AMADEOSBook}, changes to
components configurations are introduced over time, and models
need to be updated to reflect such changes.
Improving variability means anticipating certain kind
of changes and make them easier to be implemented \cite{vanGurp01}.

In the dependability \cite{Avizienis04} and performability
\cite{Meyer80} domain many works have proposed approaches to automatically
generate formal models from design models (e.g., UML models) enriched with 
information on the failure/repair processes of components.
The idea behind these works is that software and systems engineers can
take advantage of formal models without being proficient in them,
because model transformations embed the knowledge of experts in an automated
``push-a-single-button'' tool \cite{Montecchi11,Bernardi2012}.

While these approaches 
are good in providing an application-specific
abstraction to users of a certain domain, they are not flexible enough to
relieve dependability experts from the effort of modeling complex systems.
In fact, they have two main limitations:
\begin{enumerate*}[label=\roman*)]
\item they are tailored to the needs of system designers and not to those of
formal methods experts, and
\item different transformation algorithms needs to be defined for different
problems or classes of systems.
\end{enumerate*}


In this paper we address the problem of 
variability in performability models from the point of view of modeling
experts, as opposed than targeting software and systems engineers.
The focus is on specifying models considering variability, that is, anticipating
and facilitating changes.
We propose a new formalism based on \acp{SAN}
\cite{Sanders02} that we call \acp{SAN-T}.
The idea is to leave some parts of a SAN model unspecified, and to make
them depend on parameter values. Differently from what is done for example
in the Möbius tool \cite{Clark01}, in \ac{SAN-T} models the parameters can also affect the structure of the model, like the
number of cases of an activity.

This paper complements our recent work in \cite{2020TR}, in which we defined
a framework to improve reuse of performability models.
The approach is based on the concept of \emph{libraries of model
templates} that interact using well-defined interfaces and
composition rules, specified using the \ac{TMDL}.
%
The framework proposed in \cite{2020TR} is 	not tied to a specific
modeling formalism and can be applied in general to
models that are composed by superposition of state variables (i.e., ``state sharing'').
To achieve such generality, we assumed
as prerequisite the existence of i) a \emph{template-level formalism}, ii) an \emph{instance-level
formalism}, and iii) a \emph{concretize function}, which generates instance-level
models from a template-level model.
The work in this paper enables the application of the \ac{TMDL} framework with
\acp{SAN}, since we introduce here
a template-level formalism based on SANs and the associated \emph{concretize}
function.

The rest of the paper is organized as follows.
In \autoref{sec:background} we introduce the background and we discuss the
related work.
In \autoref{sec:motivating} we present the overall idea of SAN templates with a
running example, and then in \autoref{sec:sant} we give their formal definition.
In \autoref{sec:sant-instances} we define how concrete SAN models can be derived
from a SAN template (i.e., the \emph{concretize} function).
Then, in \autoref{sec:casestudy} we apply the formalism to the modeling of the
backbone network of an environment monitoring system. 
In \autoref{sec:tmdlframework} we summarize how this paper complements the
the \ac{TMDL} framework introduced in \cite{2020TR}.
Finally, conclusions are drawn in \autoref{sec:conclusions}.

\section{Background and Related Work}
\label{sec:background}

\subsection{Model-Based Evaluation}

%

Model-based evaluation \cite{Nicol04} is a well-known technique for the
verification and validation of complex systems. It consists
in estimating system-level metrics through
formal models, which typically include stochastic behavior.
Model-based evaluation plays a key role in the assessment of critical systems
and large-scale infrastructures, where exercising the real system
is not feasible or risky.

Various kind of models can be used for this task. Approaches are
typically categorized in combinatorial models and state-space models
\cite{Nicol04}. \emph{Combinatorial models} include simple formalisms
that describe which combinations of component failures lead to
system failure, e.g., \acp{FT} \cite{NASA-FT}. These models are very popular in the industry, as they are
simple to understand and they can be evaluated with well-known formulas.
However, they assume independent events and therefore they cannot
represent complex interactions between components or dynamic behavior.

On the other hand, \emph{state-space models} explicitly represent the different
states of a system and the possible transitions between them. While being more
powerful, these models can quickly become very complex, leading to
well-studied problems like state-space explosion and stiffness \cite{Nicol04}.
One of the most popular formalisms are
\acp{SPN}
and their numerous extensions \cite{Ciardo94a}.
In particular, the work in  this paper is based on \acp{SAN}, which can be
considered a variant of \acp{SPN} \cite{Sanders02}, although adopting a
different terminology (e.g., \emph{activity} instead of \emph{transition}).

\acp{SAN} are widely used
across different domains.
For example, recent work has employed models based on \acp{SAN} to
evaluate control strategies of smart grids \cite{Chiaradonna16}, the
availability of a backbone network \cite{Nencioni17}, the performance of scheduling
algorithms \cite{EntezariMaleki16}, different metrics \cite{Basile18} and
performability \cite{DaSilva19} in the railway domain, and the quality of experience of
a distributed interactive application \cite{Veeraragavan16}.

\subsection{Stochastic Activity Networks (SAN)}
\label{sec:san}

A formal definition of Stochastic Activity Networks (SANs) was given by Sanders
and Meyer in \cite{Sanders02}. We recall here the basic definitions, on
which we will base later for the definition of SAN templates.

An \ac{AN} is an eight-tuple \cite{Sanders02}:
\begin{equation}
AN = (P, A, I, O, \gamma, \tau, \iota, o),
\end{equation}
where
  $P$ is a finite set of \emph{places};
  $A$ is a finite set of \emph{activities};
  $I$ is a finite set of \emph{input gates}; and
  $O$ is a finite set of \emph{output gates}.
  The function $\gamma\colon A\to\mathbb{N}^+$ specifies
  the number of \emph{cases} for each activity, that is, the number of possible choices upon execution of that activity.
  $\tau\colon A \to \{\mathrm{\emph{timed}},\mathrm{\emph{instantaneous}}\}$
  specifies the type of each activity;
  $\iota\colon I\to A$ maps input gates to activities; and
  $o\colon O\to\{(a,c)\mid a\in A\wedge c
  \in\{1,2,\ldots,\gamma(a)\}\}$ maps output gates to cases of activities.

Similarly to Petri nets, places can hold \emph{tokens}.
The number of tokens in each places determines the state of the network, also called its marking.
More formally, if $S$ is a set of places ($S\subseteq P$), a
\emph{marking} of $S$ is a mapping $\mu\colon S\to\mathbb{N}$.
The value $\mu(p)$ is the
marking of place $p$, i.e., the number of tokens it holds.
The set of possible markings of
$S$ is the set of functions $M_S=\{\mu\mid\mu\colon S\to\mathbb{N}\}$.

An \emph{input gate} is defined as a triple
$(G,e,f)$, where $G\subseteq P$ is the set of input places associated with the gate, $e\colon
M_G\to\{0,1\}$ is the enabling predicate of the gate, and $f\colon M_G\to M_G$
is the input function of the gate.
An \emph{output gate} is a pair $(G,f)$, where $G\subseteq P$ is the set
of output places associated with the gate, and $f\colon M_G\to M_G$
is the output function.

An input gate $g=(G,e,f)$ \emph{holds} in a marking $\mu$ if $e(\mu_G)=1$.
We say that an activity $a$ is \emph{enabled} in a marking $\mu$ if all the
input gates associated with it hold.
Intuitively, the behavior of the network is regulated by the following rules:
\begin{inparaenum}[i)]
\item when an activity is enabled it can \emph{fire};
\item instantaneous activities have priority over timed activities; and
\item when an activity fires, one of its cases is selected.
\end{inparaenum}
When an activity $a$ fires in marking $\mu$, the new marking is given by
$\mu'=f_{O_n}(\ldots f_{O_{1}}(f_{I_m}(\ldots f_{I_1}(\mu))))$, where
$g_i=(G_{I_i},e_{I_i},f_{I_i})$ is the $i$-th input gate of the activity,
and $o_j=(G_{O_j},f_{O_j})$ is the $j$-th output gate of the selected case.
That is, all the functions of all the input gates are computed first, and then
all the functions of the output gates are computed.
The complete characterization of the SANs behavior can be
found in \cite{Sanders02}.

A marking in which no instantaneous activities are enabled is a \emph{stable}
marking. An activity network is \emph{stabilizing} if, essentially, there is no
marking from which it is possible to fire an infinite sequence of instantaneous
activities.

Given an \ac{AN} that is
stabilizing in some initial marking $\mu_0\in M_P$, a
\emph{Stochastic} Activity Network (SAN) is formed by defining functions $C_a$,
$F_a$, and $G_a$ for each activity $a$, where: $C_a\in C$ is a function
specifying the probability distribution of its cases; $F_a\in F$ is a function specifying the
probability distribution of its firing delay; and $G_a\in G$ is a function
that describes its reactivation markings \cite{Sanders02}.

That is:
\begin{equation}
\begin{split}
SAN =\left((P, A, I, O, \gamma, \tau, \iota, o),\mu_0,C,F,G\right).
\end{split}
\end{equation}

\acp{SAN} have an intuitive graphical notation (see \cite{Sanders02}).
Places are represented as circles, instantaneous
activities as thin bars and timed activities as thick bars. Input gates are
represented as left-pointing triangles, while output gates as right-pointing
triangles. Cases are represented as small circles next to the activity; if an
activity has only one case, it is omitted from the diagram. 
Input arcs are considered a special case of input gate, in which the predicate
$e$ holds when there is at least one token in the connected place, and the
function $f$ removes one token from that place. Similarly, an output arc is
a special case of output gate in which the function $f$ simply adds one token to
the connected place.
%

Evaluation metrics are defined using \emph{reward structures}, and
under certain conditions the stochastic process underlying a SAN has an exact solution.
If it is not the case, they can be evaluated by discrete-event
simulation.
Practical support for the \ac{SAN} formalism is provided by the M\"obius
\cite{Clark01} tool.


\subsection{Related Work}
\label{sec:related}

The problem of simplifying the construction of performability models
has been approached in different ways in the literature.

Different variants of the original \acp{PN} formalism \cite{Petri66} have been
defined, some of them enabling more compact and reusable specifications.
For example, \acp{CPN} \cite{CPNBook} allow tokens to be distinguished, by
attaching information to them. In \acp{CPN} tokens can be of different data
types, called colors. Hierarchical CPNs support modularization by means of substitution
transitions,
i.e., a transition
is replaced by a whole subnet
in a more detailed model.
\acp{SRN} \cite{Muppala94} also
contain features that allow for a compact specification of SPNs, e.g., marking
dependency, variable-cardinality arcs, priorities, etc. 

These \acp{PN} extensions fold a complex Petri net model into a
compact specification. 
They can address variability aspects to some extent, for example combining
different initial markings and marking-dependent properties.
However, variable aspects are mostly limited to the behavior and not to the
structure of the model.
Besides that, these formalisms are used to specify concrete models,
albeit complex, that can be directly analyzed.
Using the terminology in \cite{2020TR}, \acp{CPN} and \acp{SRN} are
\emph{instance-level formalisms}.
This paper focuses on
specifying \emph{template-level} models, from which different instances
can be derived.

As mentioned earlier, SANs
can also be considered a variant of SPNs \cite{Sanders02}. In their M\"obius
implementation \cite{Courtney09} they support tokens having different
datatypes, including structured datatypes.
The \emph{input gate} and \emph{output gate} primitives can be used to specify
arbitrary complex predicates for the enabling of transitions (called activities)
and for the effects of transition firings.
SANs models can be composed using the
Rep/Join state sharing formalism \cite{Sanders91};
however, which state variables are composed, and how, must be specified
manually. The M\"obius implementation of SANs permits
using variables, which however can only impact the behavior of the model and
not its structure. In this paper we define parametric (``template'') SAN models, whose
structure \emph{and} behavior can depend on parameters.



A well-established research line focuses on applying Model-Driven Engineering
(MDE) \cite{Schmidt06} techniques to automatically derive
 dependability models from UML models or similar representations, e.g., see
\cite{Bernardi2012,Cinque12,Montecchi11}.
However, such approaches typically provide an application-specific abstraction
to users of a certain domain, and then they automatically derive formal models
defined by an expert.
Instead, our approach is targeted at dependability modeling experts, and it focuses on
constructing models that could be reused across different domains or
systems.

It should be noted that existing modeling frameworks, e.g.,
M\"obius \cite{Courtney09} or CPNTools \cite{CPNBook},
provide some means for reducing the specification effort.
For example, they both allow multiple instances of a submodel to be
reused.
However, instances have identical
structure, and each of them still needs to be manually connected to the rest of
the model.

The recent work in \cite{Masetti17} proposed \ac{DARep}, an efficient
method to replicate \ac{SAN} models while still maintaining their identity, as opposed to
the traditional replica operator in which instances are
indistinguishable.
The method uses a matrix to specify dependencies across state-variables of
instances, and an efficient algorithm \cite{Masetti18} to reflect them in
the discrete-event simulator generated by M\"obius.

The work on \ac{DARep} is probably the most similar to our proposal. However, there
are two main differences between the two approaches. First, \ac{DARep} focuses
on replicating identical \ac{SAN} models, with alterations in the
state-variables shared between them. In this aspect, our approach is more
general, since i) it allows variability in the entire structure of a \ac{SAN}
model (instead than on ``interfaces'' only), and ii) it permits 
to generate \ac{SAN} instances that can be composed by either Join or Rep
operators.
Second, the main motivation behind \ac{DARep} is to improve the performance of simulation
solvers \cite{Masetti17}, while we focus on the formalization of the concept of
\ac{SAN} Template itself, and we consider solution methods out of the scope
of this paper.




\section{Motivating Example}
\label{sec:motivating}

In this section we introduce an example to motivate 
the approach we are proposing. 
We first provide an overview of the reference system, then we discuss the
challenges in modeling it using ordinary \acp{SAN}, and finally we extract a
running example to be used in the rest of the paper. 

\subsection{The BEDCS System}

The \ac{BEDCS}, also known with its Portuguese name
\emph{\acs{SBCDA} (\acl{SBCDA})}, is a large-scale
environmental monitoring infrastructure owned by the National Institute of
Space Research of Brazil (INPE) \cite{Chamon06,MattielloFrancisco12}.
Among the other applications, 
it 
supports the monitoring of
the Amazon rainforest, both in Brazil as well as in other South American countries
\cite{AmazonFund,AmazonColombia}.

\begin{figure}
\centering
\includegraphics[width=.85\linewidth]{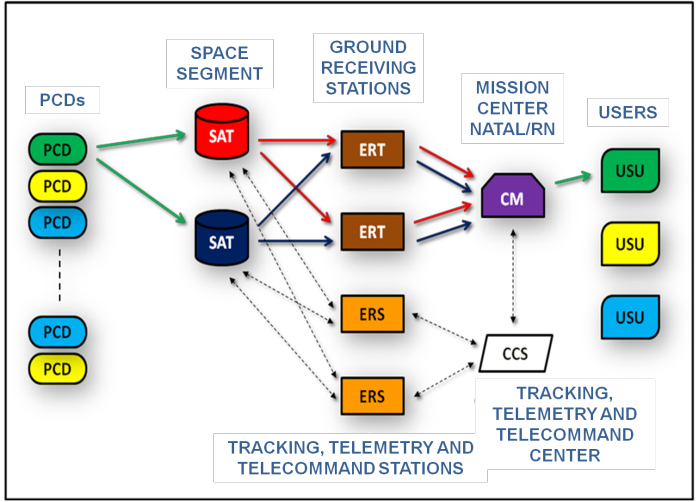}
\caption{Network architecture of the \ac{BEDCS} \cite{deCarvalho13}.}
\label{fig:bedcs-blocks}
\vspace{-4mm}
\end{figure}

The \ac{BEDCS} is composed of three segments:
space, ground, and application (\autoref{fig:bedcs-blocks}). 
The \emph{application segment} features a deployment of approximately 800
automated platforms scattered throughout the country and at sea (oceanographic buoys).
In each of these \acsp{PCD} (\emph{\acl{PCD}} --- \acl{DCP}, \acs{DCP})
groups of sensors are installed for collecting different kinds of
environmental data.
The \emph{space segment} consists of
satellites, which carry a data-collecting transponder and periodically receive
the data collected by the \acsp{DCP}.
Satellites also collect data by themselves, typically Earth images
or physics-related measurement like density of electrons.
Finally, the \emph{ground segment} consists of  ground
stations for the reception of data from satellites, a
center for remote control and tracking, and a mission center
responsible for data processing and dissemination to the end users. The data is
stored at the mission center, where they are processed and made available to end
users through a web interface.

Due to the large number of applications relying on it \cite{Fonseca14}, the
system has to fulfill strict non-functional requirements; among others:
\emph{availability} of the platform to end-users; \emph{reliability} of the
data collection functionality, in order not to miss data points;
\emph{performance} of data transmission, both \acsp{DCP} to satellites and
satellites to ground station. Note that low performance in data transmission may
also cause data loss, in case the buffers of satellites or \acsp{DCP}
become full.

The system, which is in operation since the 90's, has gained importance in the
Brazilian and International community over time, resulting in an increasing
demand for system modernization, for the provision of new services, 
and for improvements in performance and dependability.
In this context, model-based evaluation is a valuable tool to help engineers 
understanding the impact of maintenance actions, and to support informed design
decisions.
%
%
%
%


The \ac{BEDCS} is one of the two case studies of the \acs{ADVANCE} project
\cite{ADVANCE}, whose objective is to define new \acl{VV} techniques for
\aclp{CPS}.
Within the project, the system is being analyzed according to different points
of view, supporting INPE in its evolutive maintenance efforts.

\subsection{Challenges}

The \ac{BEDCS} is a representative example of a \ac{CPSoS}: it is composed of a
large number of components, with complex interactions between them that 
possibly change over time, and it is organized as \ac{SoS} architecture. That
is, its \acp{CS} may have different governance and ownership (e.g.,
satellites and sensors deployed by third-parties) and they are put together to
provide a higher goal that could not be provided by individual \acp{CS} alone.

Model-based evaluation is the primary evaluation means for this kind of system, due
to the difficulties in applying experimental approaches, especially at system
level. However, when modeling systems like \ac{BEDCS} with \acp{SAN} or similar
formalisms, practical issues concerning the scalability of these models arise.
Scalability concerning the solution process is a well-known problem that 
has been addressed in several ways in the literature; e.g., see \cite{Nicol04,Ciardo12}.
Instead, in this paper we address the challenges concerning the
\emph{specification and maintenance} of such models.

In this perspective, the main challenge is represented by the large number of
similar components, which exhibit variability aspects due for example of
different roles, location, or configuration. Furthermore, these aspects can
change over time, leading to what in software engineering is known as
\emph{variability in space and time} \cite{Seidl14}.
In the \ac{BEDCS} case, variability exists in different aspects.
As a simple example, each satellite may use a subset of the
available communication protocols, which has impact on which \acsp{DCP} it may
receive data from.
More in general, different kinds of relations and dependencies may exist between
system components. Often, a certain relation or global behavior can be described
in a general way; however, its practical \emph{instance} varies depending on
the number and kind of components involved, and possibly other
context-dependent parameters.




To some extent, this problem is being addressed by 
modularization. When modeling a
complex system with \acp{SAN}, the complete model is
typically built out of a well-defined set of submodels representing specific aspects of the
system, which are then composed by state sharing following predefined rules.
Basic building blocks are first identified (e.g., components or functions), and
then examples of the corresponding \ac{SAN} models are described.
%
Variability is addressed by defining these building blocks in a general way,
as ``templates'', explaining how variants can be derived from a general abstract
structure.

In the literature, this kind of approach has been used by different authors in
the construction of models based on \acp{SAN}, but always in an informal way.
See for example
\cite{Kanoun99,Kanoun00,Rabah03,Bondavalli09,Battista13,Chiaradonna16,Veeraragavan16,Nencioni17,Ge17}.
In fact, while a \acp{SAN} can be used to accurately describe the model of a
specific instance of the building block, with its specific characteristics, the
general structure of the model and its possible variations can only be described
with examples or with descriptions in natural language.
This is especially true when variations include modifications to the model
structure, e.g., number and names of places or number of cases of activities.

The \ac{SAN-T} formalism that we define in this paper aims to solve this
problem, by providing a formal way to specify a general (i.e., template)
\ac{SAN} model \emph{and} its possible variations.
The idea is to provide an abstract representation of multiple \ac{SAN} models
that exhibit similar structure and behavior, but having some systematic
differences that can be parameterized.
Then, from such base skeleton, different variants can be generated, based on the
values assigned to its parameters.

\subsection{Running Example: The \emph{User} Model}
\label{sec:running}

We introduce here one of the building blocks of the \ac{BEDCS} model, which will
be used as running example in the rest of the paper.

Among the other things, the \ac{BEDCS} model has to take into account for
different service kinds (e.g., raw picture data transfer or
telemetry to satellites), having different characteristics but a
similar behavior.
Also, different kinds of user models are needed, having access to different
subsets of these services, and accessing services with different
probabilities. Metrics of interest are both related to performance (e.g.,
throughput) and reliability (e.g., disconnection probability).

\newcommand{\runningexamplescale}{.55}
\begin{figure*}
    \begin{subfigure}[b]{0.32\textwidth}
		\includegraphics[scale=\runningexamplescale]{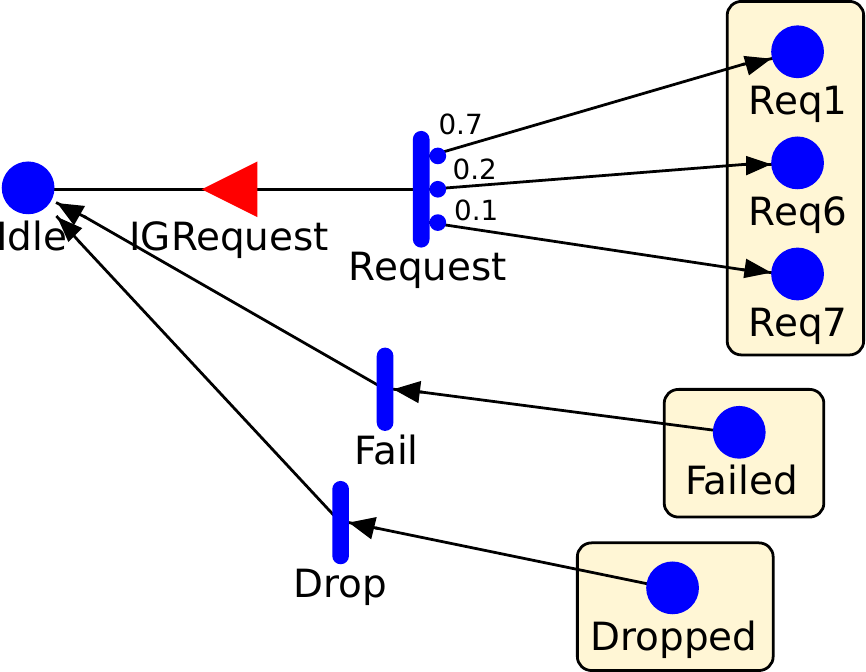}
        \caption{\emph{UserInternal} \ac{SAN} model}
        \label{fig:motivation-san1}
    \end{subfigure}
    \hfill
    \begin{subfigure}[b]{0.32\textwidth}
		\includegraphics[scale=\runningexamplescale]{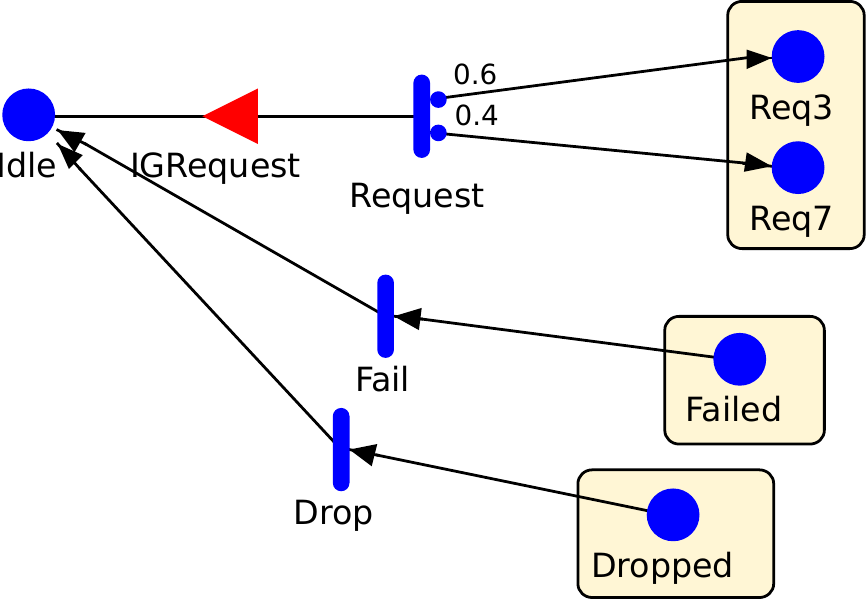}
        \caption{\emph{UserPress} \ac{SAN} model}
        \label{fig:motivation-san2}
    \end{subfigure}
    \hfill
    \begin{subfigure}[b]{0.32\textwidth}
		\includegraphics[scale=\runningexamplescale]{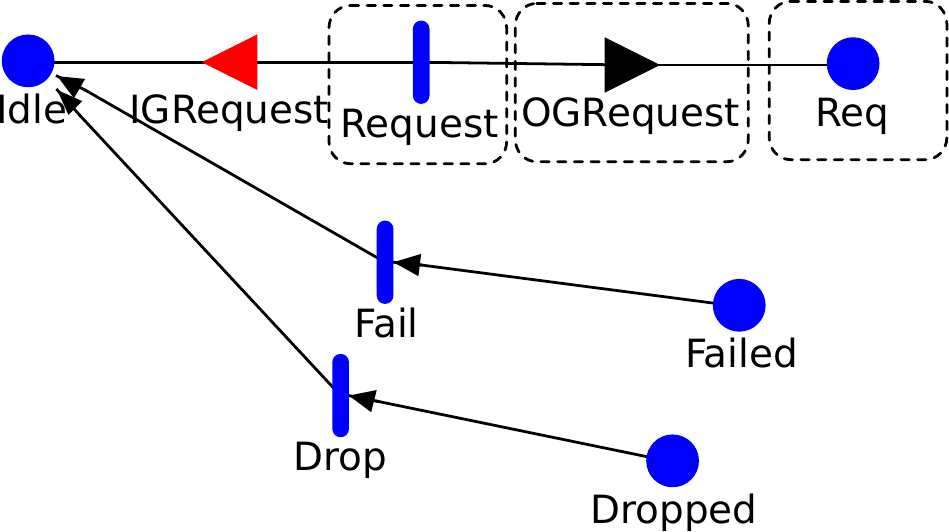}
		\caption{\emph{User} \ac{SAN-T} model}
		\label{fig:motivation-sant}
	\end{subfigure}
    \caption{The general idea of SAN templates. The two SAN models in \subref{fig:motivation-san1}
    and \subref{fig:motivation-san2} have a similar structure and behavior, and are meant to be connected with
    other SAN models in the same way, that is, sharing the boxed places. This structure can be abstracted into 
    a \sant{} model as the one depicted in \subref{fig:motivation-sant}, where dashed elements represent template
    versions of SAN elements, i.e., activity templates and place templates.}
    \label{fig:sant-general-idea}
    \vspace{-3mm}
\end{figure*}

The basic idea behind \acp{SAN-T} is visualized in
\autoref{fig:sant-general-idea}.
The \acp{SAN} models in Figures
 \ref{fig:motivation-san1}--\subref{fig:motivation-san2} are adapted
from \cite{Bondavalli09}, in which we modeled a
vehicular network. We chose this example for its simplicity; a
more extensive application of the \ac{SAN-T} formalism is
described later in \autoref{sec:casestudy}.



The behavior represented by the two SAN models in Figures
\ref{fig:motivation-san1}--\subref{fig:motivation-san2} is the following.
Each user is initially in idle state, and may then request a network
service. With a certain probability they can request one of the services that
are available to them, by adding a token in the corresponding place.
While the service is being delivered, a token stays in the place with the
corresponding identifier (e.g., \texttt{Req1} or \texttt{Req6}).
The request can fail or be dropped; in these cases a token is received in
the corresponding place, and the user returns to idle state. 
%
The figure shows two
instances of the model of an user:
an \emph{internal} user may request services 1, 6, and 7,
with the respective probabilities, while a \emph{press} user may request
services 3 and 7 only, with different probabilities.

It is clear that the two models 
have a similar
structure. In fact, they differ only by: i) the number of services available
to the user, ii) the identifiers of those services, and iii)
the probabilities of the user requesting each service.
The structure of these two models can be generalized by establishing  the
following informal rule:
\emph{``Create one place ReqX for each of the services that are available to
the user, and name them according to the identifier of these services. The
activity Request should have the same amount of cases as the number of ReqX
places, and each of them should have an output arc connecting the case to the
corresponding ReqX place.''}
This would result in a SAN ``template'', depicted in
\autoref{fig:motivation-sant}, which abstracts the common
structure among models of different users.


It should be noted that the template in \autoref{fig:motivation-sant} does not
represent the instances in Figures
\ref{fig:motivation-san1}--\subref{fig:motivation-san2} only, but in general any
\ac{SAN} model that follows the same pattern.
Having to maintain similar models that only differ from some details is a
common issue in the modeling of complex systems. 
The actual formal definition of the \emph{User} \sant{} model discussed above is provided later in \autoref{sec:sant-user}.

\section{Stochastic Activity Network Templates}
\label{sec:sant}

\subsection{Preliminary Definitions}
\label{sec:preliminary-definitions}

We first introduce some basic notations that will be used in the rest
of the paper. In particular, the following definitions clarify what is a
parameter of a template model, and how it connects to the rest of the formalism.
 
We adopt the definitions of \emph{sort},
\emph{operator}, \emph{term} and \emph{assignment} from the ISO/IEC 15909
standard \cite{ISO15909-1}, which apply to a wide range of \ac{PN}-based
formalisms, including \acp{SAN}. 
According to this formalization, the set of possible values held by a place
is defined by its associated \emph{sort} (i.e., type).

A \emph{many-sorted signature} is a pair $(S,O)$, where
$S$ is a set of sorts and $O$ is a set of \emph{operators}, together with their
arity.
%
The arity is a function $O\to S^*\times S$, where
$S^*$ is the set of finite sequences over $S$, including the empty string
$\varepsilon$.
The arity function defines, for each operator, the number and sort of its input
parameters ($S^*$), and the sort of the produced result ($S$).
An operator can be denoted as $o_{(\sigma,s)}$,
where $\sigma\in S^*$ are the input sorts, and $s\in S$ is the output sort.
Constants are operators with empty input sorts, and are denoted as
$o_{(\varepsilon,s)}$ or simply $o_s$.

We denote with 
$\Delta$ a set of parameters; an element of $\Delta$ of sort
$s\in S$ is denoted with $\delta_s$. 
$\Delta_s\subseteq\Delta$ is therefore the set of parameters of sort $s$.

\emph{Terms} of sort $s\in S$ may be built from a signature $(S,O)$ and a
set of parameters $\Delta$. 
Intuitively, these are all the possible expressions of sort $s$
made of any legit combination of operators in $O$ and parameters in $\Delta$
\cite{ISO15909-1}. 
The set of terms of sort $s$ is denoted by
$\tterms{s}$. 
To simplify the notation, in the rest of the paper we will
use $\ttermssimple[s]$, unless there are ambiguities on the adopted $O$ and
$\Delta$ sets.

A \emph{many-sorted algebra} $H=(S_H,O_H)$ provides an interpretation of a
signature $(S,O)$. For every sort $s\in S$ there is a corresponding set of values $H_s\in S_H$, and for every
operator $o_{(s_1\ldots s_n,s)}\in O$ there is a corresponding function in
$o_H\in O_H$, such that $o_H\colon H_{s_1}\times\ldots\times H_{s_n}\to H_s$. 

Given a many-sorted algebra $H$, and many-sorted parameters in $\Delta$, an
\emph{assignment} for $\Delta$ under $H$ is a family of functions $\xi$,
comprising a function $\xi_s\colon\Delta_s\to H_s$ for each sort $s\in S$.
The concept of assignment may be extended to terms, thus obtaining the family of
functions $\mathit{Val}_\xi$ comprising the function
$\mathit{Val}_{s,\xi}\colon\ttermssimple\to H_s$ for each sort $s\in S$ 
\cite{ISO15909-1}.

To support the subsequent definitions, we require the existence of at least the
``{integer}'', ``{real}'', ``{boolean}'', ``{set of integers}'', and ``{set of
reals}'' sorts, in which we consider sets to be ordered. Formally,
we assume a signature $({S},{O})$, such that  
$\{\tint,\treal,\tbool,\tintset,\trealset\}\subseteq
{S}$, and ${O}$ contains the common operators applicable on such sorts.
In particular, besides the standard arithmetic operators, in the rest of the
paper we will use the \emph{size} operator, $|x|$, which returns the
number of elements in a set $x$, and the \emph{element at} operator,
$x[i]$, which returns the $i$-th element in a set $x$.
The corresponding many-sorted algebra is $({S}_H,O_H)$, with
$\{\mathbb{N},\mathbb{R},\{0,1\},\mathcal{P}(\mathbb{N}),\mathcal{P}(\mathbb{R})\}\subseteq
{S}_H$, and ${O}_H$ containing the set of functions corresponding to operators
in $O$, which are not detailed here for simplicity.

\subsection{\ac{SAN-T} Formal Definition}
\label{sec:sant-formal}



Based on the previous definitions, we can now introduce the formal definition of
\ac{SAN-T}.
Formally, a \emph{\acl{SAN-T}} (\ac{SAN-T}) is a
tuple:
\begin{equation}
\label{eq:sant}
{SAN\text{-}T} = (\Delta,\tilde{P},\tilde{A},\tilde{I},\tilde{O},
\tilde{\gamma}, \tilde{\tau}, \tilde{\iota},\tilde{o}, \tilde\mu_0,\tilde
C,\tilde F,\tilde G),
\end{equation}
 where
$\Delta$ is a set of parameters, and elements marked
with a tilde accent, $\tilde{\cdot}$, are modified versions of elements
existing in plain \acp{SAN} (see \autoref{sec:san}), reformulated to take
parameters into account.

In more details:
\begin{itemize}
  \item $\Delta$ is the sorted set of {\bf\em parameters} of the template.
  \item $\tilde{P}$ is a finite set of {\bf\em place templates}. A place
  template can be seen as a placeholder for multiple places that, in a regular SAN model,
  would be strongly related to each other and would vary in different instances
  of the same template.
  Based on parameters' values, a template place will be expanded to a precise
  set of concrete places. Place \texttt{Req} in \autoref{fig:motivation-sant} is
  an example of place template.

  Formally, a place template is defined as a pair $(\tau,k)$, where $\tau$ is
  the name of the place, and $k\in\ttermssimple[\tintset]$ is its multiplicity.
  Evaluating the term $k$ with respect to an assignment $\xi$ identifies a set
  of integer indices $K\subset\mathbb{N}$.
  Such indices determine the set of places to which, with the given assignment
  of parameters, the place template is expanded.
  Normal places (i.e., those always expanding to a single place of ordinary
  SANs) are those for which $\mathit{Val}_\xi(k)=\{1\}$ for any assignment $\xi$.
  \item $\tilde{A}$ is a finite set of {\bf\em activity templates}.
  \item $\tilde{I}$ is a finite set of {\bf\em input gate templates}.
  \item $\tilde{O}$ is a finite set of {\bf\em output gate templates}.
  \item $\tilde{\gamma}\colon \tilde{A}\to\ttermssimple[\tint]$ specifies the
  {\bf\em number of cases} for each activity template.
  For any activity template $\tilde{a}\in\tilde{A}$, evaluating
  $\tilde{\gamma}(\tilde{a})$ with respect to an assignment $\xi$ yields an
  integer number, which determines the number of cases of $\tilde{a}$
  under that assignment, i.e.,
  $\mathit{Val}_\xi(\tilde{\gamma}(\tilde{a}))\in\mathbb{N}$.
  \item $\tilde{\tau}\colon \tilde{A} \to
  \{\mathrm{\emph{timed}},\mathrm{\emph{instantaneous}}\}$
  specifies the {\bf\em kind of each activity template}, exactly as in ordinary
  \acp{SAN}.
  \item $\tilde{\iota}\colon \tilde{I}\to \tilde{A}$ maps {\bf\em input gate
  templates to activity templates}.
  \item $\tilde{o}\colon \tilde{O}\to\tilde{A}$ maps {\bf\em output gate
  templates to activity templates}.
\end{itemize}


In order to completely define the elements of a \ac{SAN-T}, the concept of marking needs to be extended.
In particular, we need to take into account for the existence of place
templates.
In general, the marking of a place template is not a single
value (like in ordinary \acp{SAN}), but a \emph{function}, which associates a
value to each index of the place template. We call this function the {\bf\em
marking template} of a place.

Let $F_{\mathbb{N}}$ be the set of all the possible functions 
$f\colon\mathbb{N}\to\mathbb{N}$, i.e.,
$F_{\mathbb{N}}=\{f\mid f\colon\mathbb{N}\to\mathbb{N}\}$.
If $\tilde{S}\subseteq\tilde{P}$ is a set
of place templates, a {\bf\em marking} of $\tilde{S}$ is a
mapping $\tilde\mu\colon \tilde{S}\to F_\mathbb{N}$.
In the particular case in which a place does not have variability, the marking
template will be a constant function.
%
The set of possible markings of $\tilde S$ is the set of functions
$\tilde M_{\tilde S}=\{\tilde\mu\mid\tilde\mu\colon
\tilde S\to F_\mathbb{N}\}$.


Based on these definitions of \emph{marking} and \emph{marking template}, most
of the other elements of a \ac{SAN-T} can be defined by adapting the definitions in \cite{Sanders02} to
made dependent on the assignment of parameters $\xi$.
In the following, we denote with $\Xi$ the set of all the possible assignments.

As in ordinary \acp{SAN}, an \emph{input gate template} defines an enabling condition
for an activity template, and an input function that specifies how the marking
is altered by the firing of the activity. An input gate template will always
result in a single input gate in the concrete \ac{SAN} model. Still, the
projected output gate may depend on the assignment of parameters.
Formally, an \emph{input gate template}
is defined as
a triple $(\tilde{G},\tilde{e},\tilde{f})$, where
 $\tilde G\subseteq \tilde{P}$ is the
set of input places associated with the gate, $\tilde{e}\colon
\tilde M_{\tilde G}\times\Xi\to\{\mathit{true},\mathit{false}\}$ is the enabling
predicate, and $\tilde f\colon \tilde M_{\tilde
G}\times\Xi\to \tilde M_{\tilde G}$ is the input function.


In ordinary \acp{SAN}, an output gate defines an output function that is
executed upon the firing of an activity. Differently from an input gate, it is
associated to individual cases of an activity.
In \acp{SAN-T}, an \emph{output gate template} has a similar purpose. However,
since the number of cases of an activity template is not known beforehand, the
gate is connected directly to the activity. When a regular \ac{SAN} is
generated from the template, an output gate template will be expanded to multiple concrete output gates, depending on
the number of cases of the activity to which it is connected.

Formally, an \emph{output gate template} is a
pair $(\tilde G,\tilde f)$, where $\tilde G\subseteq\tilde P$ is the set of
output places associated with the gate, and
$\tilde f \colon \tilde M_{\tilde
G}\times\mathbb{N}\times\Xi\to \tilde M_{\tilde G}$
is the output function of the gate.
It should be noted that the output function $\tilde f$ depends on the index of
the case of the associated activity template ($\mathbb{N}$), as well as on
the assignment of values to parameters ($\Xi$).


The probability of cases of an activity template is given by the \emph{case
distribution assignment} $\tilde C$, which defines a function $\tilde C_{\tilde
a}\in\tilde{C}$ for each activity template $\tilde a\in\tilde A$.
Such functions also depend on parameters,
  thus $\tilde C_{\tilde a}\colon \tilde M_{\tilde{P}(\tilde a)}\times \mathbb{N}^+\times\Xi \to[0,1]$, where
  $\tilde{P}(\tilde a)$ is the set of input and output places
  of the activity. For the model to be well-formed,
  $\tilde C_{\tilde a}(\mu,i,\xi)=0$ should hold
  $\forall i>\mathit{Val}_\xi(\tilde\gamma(\tilde a))$, i.e., the
  probability of cases beyond those generated with the 
  given assignment $\xi$ should be zero.

Similarly, the firing time of activities is given by 
  the \emph{activity time distribution assignment} $\tilde F$,
 which defines a function $\tilde F_a\in\tilde F$
 for any timed activity template $a$, with $\tilde F_a\colon\mathbb{R}\times
  \tilde M_{\tilde P}\times\Xi\to[0,1]$. That is, the probability of a certain
  firing time ($\mathbb{R}$) depends on the marking ($\tilde M_{\tilde P}$) and
  on the parameters assignment ($\Xi$).

The reactivation function of activity templates is given by 
  the reactivation function assignment $\tilde G$, such that for any timed
  activity template $a$, function $\tilde G_a\in\tilde G$ defines the
  reactivation markings, with $\tilde G_a\colon \tilde M_{\tilde P}\times\Xi\to \wp(\tilde
  M_{\tilde P})$ and 
  $\wp(\tilde M_{\tilde P})$ denoting the power set of $\tilde M_{\tilde P}$.

Finally, the initial marking of a \ac{SAN-T} should also depend on the
assignment of values to parameters.
For this reason, it is defined by the function 
 $\tilde\mu_0\colon\Xi\rightarrow\tilde M_{\tilde P}$.
It should be noted that
the original definition of \acp{SAN} requires the initial
marking $\mu_0(\xi)\in\tilde M_{\tilde P}$ to be a stable
marking in which the network is stabilizing (see \autoref{sec:san}).
However, because in \ac{SAN-T} the actual structure of the model is not
completely specified until a value is assigned to all the parameters, we
relax this constraint.
Well-formedness checks on the structure of the resulting \ac{SAN} models can be
performed at the time of instantiation, based on existing techniques that are
applied to ordinary SAN models (e.g., \cite{Deavours99}).

\subsection{Arc Templates}
\label{sec:arc-templates}

One of the distinguishing features of \acp{SPN} and their extensions is their
convenient graphical notation, which permits describing most aspect of a model
using a diagram. 
In particular, \acp{SAN} use {input arcs} and {output arcs}, represented by arrows in the
diagram, as the graphical representation of particular cases of input gates and output gates,
respectively.
Following the same idea, we propose a graphical representation of a subset of
the new concepts introduced in \acp{SAN-T}.

\subsubsection{Output Arc Templates}

In ordinary \acp{SAN}, an \emph{output arc} connecting activity $a$ and place
$p$ represents an output gate whose function simply adds a token to place $p$. 

To use a similar notation in \acp{SAN-T} we need to take into
account for the varying number of cases of the activity, and for the
multiplicity of the place. 
Without additional information, the interpretation of a ``normal'' output arc
is ambiguous. For example, it is not clear whether the arc should add one token
to all the instances of the place template, or only to one of them.
Therefore, we extend this concept introducing output arc templates, which can be used in \acs{SAN-T}
models.

Considering a \ac{SAN-T} model as defined above, an \emph{output arc template}
connects an activity template ${a}_{\text{src}}\in\tilde{A}$ to a place template
$p_\text{dest}\in\tilde{P}$. An output arc template has a \emph{label} that
defines the function $f$ of the corresponding gate template in a compact way.
The syntax of the label is given by the following grammar.

\vspace{-2mm}
\begin{small}
\begin{bnf*}
  \bnfprod{oat-label}
    {\bnfpn{out} \bnfor \bnfpn{int}\bnfts{$\to$}\bnfpn{out} \bnfor
    \bnfpn{int}\bnfts{$\to$}\bnfpn{out} \bnfts{/}
    \bnfpn{out}} \\
  \bnfprod{out}
    { \bnfpn{int} \bnfor \bnfts{$+$} \bnfpn{int} }\\
  \bnfprod{int}
    {\bnftd{any integer term $t\in\tterms[\Delta\cup\{\otimes,\odot\}]{\tint}$}}\\
\end{bnf*}
\end{small}
\vspace{-5mm}

%
Here the term $\odot$ is a placeholder for the index of the case of the
associated activity template ${a}_\text{src}$, while $\otimes$ is a placeholder for
the index of the concrete places generated from place template
${p}_\text{dest}$.

The semantics of such label expressions is as follows. A label may specify an
unconditional expression, $\bnfpn{out}$, or a conditional
expression, $\bnfpn{int}\to\bnfpn{out}/\bnfpn{out}$.

When an \emph{unconditional expression} is specified, it means that the same
expression is used for all the places derived from place template
${p}_\text{dest}$.
The expression may specify that the marking of the place(s) should be
\emph{set} to a certain value, $\bnfpn{int}$, or that a number
of tokens should be \emph{added} to the marking, $+\bnfpn{int}$.
The actual value is specified by a term (i.e., expression) of integer type.
Note that in this case the integer terms may also include the $\odot$ and $\otimes$ operators, and may therefore depend on them. For example, the label ``$+3\otimes$'' specifies that to each place
derived from ${p}_\text{dest}$ it should be added a number of tokens equal to three times its
index.

A \emph{conditional expression} allows specifying a different expression for a
specific instance of the place template. The expression
$\bnfpn{int}\to\bnfpn{out}$ means that for the place template with index
$\bnfpn{int}$ the specification $\bnfpn{out}$ will be used, while for 
all the other ones the marking is left unchanged.
An explicit assignment can be added for the other places, instead of leaving
them unchanged, by adding a second $\bnfpn{out}$ element.
For example, the label ``$1\to +2/ 0$'' means that for instance of
place ${p_\text{dest}}$ having index $1$ the marking should be incremented by
two tokens, while for all the other instances it should be set to zero.

When no label is specified, the label ``$+1$'' is assumed, that is, one
token is added to all the instances of the $p_\text{dest}$ place.
Note that when the place template has multiplicity $\{1\}$ and the
number of cases of the transition is fixed, an arc with label ``$+1$''
corresponds to a ``normal'' output arc.

\subsubsection{Input Arc Templates}

A similar approach can be followed for input arcs. However, in this case the
label must also specify the input predicate, in addition to the input function.

An \emph{input arc template}
connects a place template $p_\text{src}\in\tilde{P}$
to an activity template ${a}_{\text{dest}}\in\tilde{A}$. 
The \emph{label} of an input arc template is defined by the following grammar,
where $\otimes$ is a placeholder for the index of places derived from
$p_\text{src}$.

\vspace{-1mm}
\begin{small}
\begin{bnf*}
  \bnfprod{iat-label}
    { \bnfts{[} \bnfpn{pred} \bnfts{] } \bnfpn{func} \bnfor
    \bnfts{$-$}\bnfpn{int}  }\\
  \bnfprod{pred}
    { \bnfts{$\forall$} \bnfpn{cond}
      \bnfor \bnfts{$\exists$} \bnfpn{cond} 
      \bnfor \bnfpn{int}\bnfpn{cond}
    }\\
  \bnfprod{cond}
    { \bnfts{=} \bnfpn{int}
      \bnfor \bnfts{$>$} \bnfpn{int} 
      \bnfor \bnfts{$\geq$}\bnfpn{int}
    }\\
  \bnfprod{func}
    { \bnfpn{int} \bnfor \bnfts{$-$} \bnfpn{int} }\\
  \bnfprod{int}
    {\bnftd{any integer term $t\in\tterms[\Delta\cup\{\otimes\}]{\tint}$}}\\
\end{bnf*}
\end{small}
\vspace{-5mm}

The label of an input gate template may specify explicitly both the predicate
and the input function, $[\bnfpn{pred}]\bnfpn{func}$,
or specify the input function only,
$-\bnfpn{int}$, and leave the predicate implicit.
The input function $\bnfpn{func}$ is specified in a similar way as for output
arc templates:
it can set the marking of all the places to a specific value, or it can subtract a
certain number of tokens.

The predicate is composed of a quantifier and a condition.
The condition $\bnfpn{cond}$ specifies a condition on the marking of the
connected place template, while the quantifier specifies on which instances of
that place the condition should hold.

The predicate $\forall\bnfpn{cond}$ is true 
when the markings of all the concrete places derived from
$p_\text{src}$ satisfy the condition $\bnfpn{cond}$.
The predicate $\exists\bnfpn{cond}$ is true if at
least one place instance satisfies that condition. Finally, the predicate
$\bnfpn{int}\bnfpn{cond}$ is true if the marking of the place instance having
index corresponding to $\bnfpn{int}$ satisfies the condition.
When the predicate is stated explicitly, the input function is applied only to
places satisfying the condition, that is: all the places in the case of
$\forall\bnfpn{cond}$; only the places satisfying the
condition in the case of $\exists\bnfpn{cond}$; and only the place with index
$\bnfpn{int}$ in the case of $\bnfpn{int}\bnfpn{cond}$.

For example, the label ``$[\exists=1]\;0$'' means
that:
i) the predicate is true if at least one of the place instances
has exactly one token, and ii) the function sets the marking of all the
place instances satisfying the condition to zero.

When no $\bnfpn{pred}$ term is specified, the input function may only remove
a certain number of tokens, and the predicate is considered to hold if all the
place instances contain at least that number of tokens.
If no label is specified at all, ``$-1$'' is assumed, that is, the gate is
enabled if all the instances of the place contain at least one token, and the
function removes one token from all of them. Note that, similarly as for output
arc templates, an arc with label ``$-1$'' connected to place template
with multiplicity $\{1\}$ corresponds to a ``normal'' input arc.

\subsection{Graphical Notation}

Besides the extended notation for input arc templates and output arc templates,
we also adopt some conventions in the graphical representation of \acp{SAN-T}
models.

The main difference in notation is that ``template elements'' in the model
are surrounded with a dashed line, as previously shown in
\autoref{fig:motivation-sant}. 
With \emph{template elements}, we mean model elements
that have some kind of variability, in particular:
\begin{inparaenum}[i)]
\item place templates having non-unary multiplicity (i.e., $k\neq\{1\}$),
\item activities with a variable number of cases (i.e., $\tilde\gamma(\tilde a)$ is not constant),
\item input gate templates connected to a place template with non-unary
multiplicity, 
\item output gate templates connected to a place template with non-unary
multiplicity, and
\item output gate templates connected to an activity having a variable number of
cases.
\end{inparaenum}

Highlighting elements that have variability helps the modeler to better
understand which parts of the model will change in the concrete \acp{SAN}
instances, that is, which parts of the model carry some variability aspect. 


\subsection{Example: \emph{User} \ac{SAN-T} Model}
\label{sec:sant-user}

We now apply the proposed formulation of \acp{SAN-T} to the running
example introduced in \autoref{sec:running}.
%
A \ac{SAN-T} model that abstracts the behavior of a generic \emph{User}
model has been shown, informally, in
\autoref{fig:motivation-sant}.
Here we show how such \ac{SAN} Template can be specified in a formal way,
according to the definitions given in the previous section.

Two parameters can be identified for the \texttt{User} template model. The
first, $s$, identifies the number and indices of services that the user can
access, and it is therefore of type ``set of integers''. The second,
$pb$, determines the probabilities of being selected of the different
services, and it is of type ``set of reals''.

The variable elements of the model are essentially the activity template
\texttt{Request}, its associated output gate template \texttt{OGRequest}, and
the place template \texttt{Req} (see \autoref{fig:motivation-sant}).
The \texttt{Request} activity has a variable number of cases, given
by the cardinality of the array of integers assigned to parameter $s$, and each
of these cases is selected with a probability given by parameter $pb$. Place
template \texttt{Req} is expanded to a number of concrete places that is again
given by the cardinality of $s$. The selection of case $i$ of
the \texttt{Request} activity template results in the addition of a token in place
\texttt{Req}$_i$.

\begin{figure}
   	\centering
	\includegraphics[scale=\runningexamplescale]{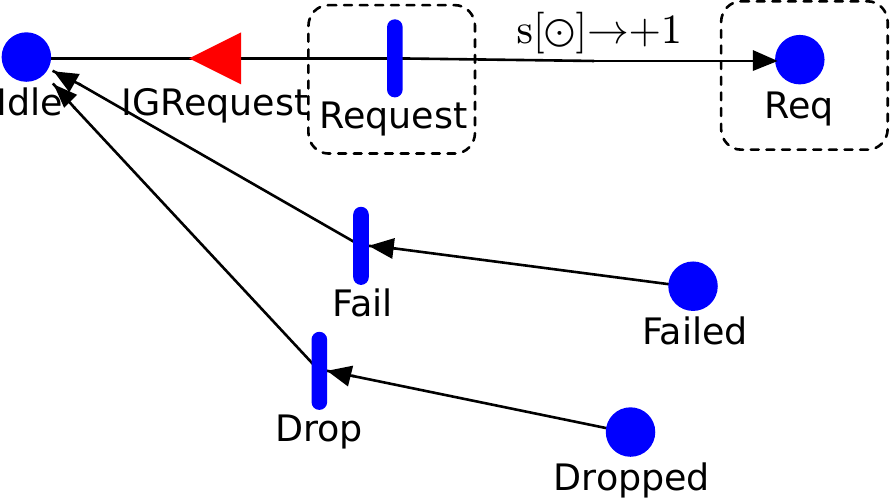}
	\caption{\emph{User} \sant{} model with the proposed notation for output arc templates.}
	\label{fig:sant-newnotation}
	\vspace{-4mm}
\end{figure}

The graphical representation of the model, using the proposed
notation, is shown in \autoref{fig:sant-newnotation}. The output arc template
connecting activity \texttt{Request} and place \texttt{Req}
has the label ``$s[\odot]\to +1$'', which can be interpreted as follows:
``{\em The $i$-th concrete output gate adds one token to the instance of
the Req place having index $s[i]$, and leaves the other
places unchanged}.'' The complete formal specification of the \emph{User} SAN-T
model is provided in the following.

As a support to the specification in this section, and to those provided later,
we define the following functions in $F_\mathbb{N}$, which can thus be used to
define the marking of the SAN-T (see \autoref{sec:sant-formal}).

\begin{small}
\begin{align*}
\label{eq:marking-templates}
f^k(n)&=k,\quad \forall n\in\mathbb{N}; \\
f^k_j(n)&=\left\{
\begin{array}{cl}\stepcounter{equation}\tag{\theequation}
k&\text{if }n=j,\\ n&\text{otherwise};
\end{array}
\right.\\
f^k_J(n)&=\left\{
\begin{array}{cl}
k&\text{if }n\in J,\\ n&\text{otherwise},
\end{array}
\right.\quad J\subseteq\mathbb{N}.
\end{align*}
\end{small}
%
%
%
The formal specification of the \emph{User} SAN-T model is then as follows.

\begin{small}
\begin{align*}
\label{eq:santfull}
\text{SAN-T}_{\mathtt{User}} &= (\Delta,\tilde{P},\tilde{A},\tilde{I},\tilde{O},
\tilde{\gamma}, \tilde{\tau}, \tilde{\iota},\tilde{o}, \tilde\mu_0,\tilde
C,\tilde F,\tilde G)\\
\Delta&=\{s_{\tintset},{pb}_{\trealset}\}\\
\tilde{P}&=\{(\mathtt{Idle},1),(\mathtt{Req},s),(\mathtt{Dropped},1),(\mathtt{Failed},1)\}\\
\tilde{A}&=\{\mathtt{Request},\mathtt{Fail},\mathtt{Drop}\}\\
\tilde{I}&=\{\mathtt{IGRequest},\mathtt{ArcInFail},\mathtt{ArcInDrop}\}\\
\tilde{O}&=\{\mathtt{OGRequest},\mathtt{ArcOutFail},\mathtt{ArcOutDrop}\}\\
\tilde{\gamma} &= \{ (\mathtt{Request},|s|), (\mathtt{Fail},1),
(\mathtt{Drop},1)\}\\
\tilde{\tau} &= \{ (\mathtt{Request},\mathrm{\emph{timed}}),
(\mathtt{Fail},\mathrm{\emph{instantaneous}}),\\
&\phantom{=\ \{}
(\mathtt{Drop},\mathrm{\emph{instantaneous}})\}\\
\tilde{\iota}&=\{(\mathtt{IGRequest},\mathtt{Request}),
(\mathtt{ArcInFail},\mathtt{Fail}),\\
&\phantom{=\ \{}(\mathtt{ArcInDrop},\mathtt{Drop})\}\\
\tilde{o}&=\{(\mathtt{OGRequest},\mathtt{Request}),
(\mathtt{ArcOutFail},\mathtt{Fail}),\\
&\phantom{=\ \{}(\mathtt{ArcOutDrop},\mathtt{Drop})\}\\
\tilde{\mu}_0(\xi)&=\mu'(\tilde{p})\;\forall\xi\mid\mu'(\tilde{p})=\left\{
	\begin{array}{ll}
	f^1 & \textrm{if }\tilde{p}=\mathtt{Idle},\\
	f^0 & \textrm{otherwise}.
	\end{array}
\right.\\
\tilde{C}&=\{\tilde{C}_{\mathtt{Request}},\tilde{C}_{\mathtt{Drop}},\tilde{C}_{\mathtt{Fail}}\}\stepcounter{equation}\tag{\theequation}\\
\tilde{C}_{\mathtt{Request}}(\tilde{\mu},i,\xi)&=
\left\{
	\begin{array}{cl}
	\mathit{Val}_\xi({pb}_i) & \text{if } 1\leq i \leq |s|,\\
	0 & \mathrm{otherwise}.
	\end{array}
\right.\\
\tilde{C}_{\mathtt{Drop}}(\tilde{\mu},i,\xi)&=
\tilde{C}_{\mathtt{Fail}}(\tilde{\mu},i,\xi)=
\left\{
	\begin{array}{ll}
	1 & \text{if } i=1,\\
	0 & \mathrm{otherwise}.
	\end{array}
\right.\\
\tilde{F}&=\{\tilde{F}_{\mathtt{Request}}\},\qquad
\tilde{G}=\{\tilde{G}_{\mathtt{Request}},\tilde{G}_{\mathtt{Drop}},\tilde{G}_{\mathtt{Fail}}\}.\\
\end{align*}
\end{small}
Some aspects should be highlighted to better understand the above
specification.
All the place templates have multiplicity 1, except for place
\texttt{Req}; similarly, all the activities have a fixed number of cases except
for activity \texttt{Request}.
The initial marking is 1 (actually, the function $f^1$) for place template
\texttt{Idle}, and 0 (actually, the function $f^0$) for all the others.

The set of input gate templates, $\tilde{I}$, also contains the input arcs of
the model, which are considered a special case of input gate. In particular, 
\texttt{ArcInFail} and \texttt{ArcInDrop} are the input arcs connected to the
\texttt{Fail} and \texttt{Drop} activities, respectively. Similarly, the set of
output gate templates, $\tilde{O}$, also contains the output arcs
\texttt{ArcOutFail} and \texttt{ArcOutDrop}.

The case distribution function assigns probability 1 to the first and only case
of activity templates \texttt{Drop} and \texttt{Fail}, while for the cases of
the \texttt{Req} activity template the probability of the $i$-th case is
given by the $i$-th value of parameter $p$, which is of type ``set of reals''.
For simplicity, the elements $\tilde{F}$ and $\tilde{G}$ have not been
detailed in the above specification, and they are discussed in the following.

In the reference model \cite{Bondavalli09}, the firing time of the
\texttt{Request} activity is regulated by an \emph{uniform} distribution, thus $\tilde{F}_{\mathtt{Request}}$ is set
accordingly. 
None of the activities are reactivating, that is,
$\tilde{G}_{\mathtt{Request}}=\tilde{G}_{\mathtt{Fail}}=\tilde{G}_{\mathtt{Drop}}=\emptyset$,
or, in other words, the set of reactivating markings is empty.

To fully specify the \ac{SAN-T}, we need to complete the specification of
input gates templates and output gates templates. We give here the
specification of \texttt{OGrequest} only, which is the only output gate template
in the \emph{User} template that includes variability:
\begin{equation}
\begin{split}
\tilde{G}_{\mathtt{OGRequest}}&=\{\mathtt{Req}\},\\
\tilde{f}_{\mathtt{OGRequest}}(\tilde{\mu},i,\xi)&=\mu'(\tilde{p})\;\forall\tilde
p\in\tilde P\mid\\
\mu'(\tilde{p})&= \left\{
	\begin{array}{ll}
	f^1_{s[i]}\left(\tilde{\mu}(\tilde{p})\right)  & \textrm{if
	}\tilde{p}=\mathtt{Req},\\
	\tilde{\mu}(\tilde{p}) & \textrm{otherwise}.
	\end{array}
\right.\\
\end{split}
\end{equation}
Summarizing, the above specification means that the function of 
the output gate associated with the $i$-th case should put one token into the
instance of the $\texttt{Req}$ having index $s[i]$ and remove all the tokens
in the other instances ($f^1_{s[i]}$), and leave the other places unchanged.


\section{Generation of \ac{SAN-T} Instances}
\label{sec:sant-instances}
\label{sec:concretization}

To actually use \ac{SAN-T} models for model-based evaluation, concrete
instances must be generated, by assigning values to their parameters.
Such instances, which are ordinary \ac{SAN} models, can be evaluated in
isolation, or composed into larger models using the \ac{TMDL} framework
\cite{2020TR} or the plain Rep/Join formalism \cite{Sanders91}.

\subsection{Overview}
Instances are generated by the
\emph{concretize} function, described in the following,
which generates an
ordinary SAN model from a from a pair $(S_\Delta,\xi)$.
That is, it generates a \ac{SAN} model from a
\ac{SAN-T} model $S_\Delta$ and an assignment of values to its parameters $\xi$. 


Given a SAN-T $S_\Delta$:
\begin{equation}
S_\Delta = (\Delta,\tilde{P},\tilde{A},\tilde{I},\tilde{O},
\tilde{\gamma},\tilde{\tau},\tilde{\iota},\tilde{o},\tilde{\mu}_0,\tilde{C},\tilde{F},\tilde{G}),
\end{equation}
and a parameter assignment function $\xi$, the \emph{concretize} function
generates a SAN model $S^\xi$:
\begin{equation}
\label{eq:sant-derivation}
S^\xi = (P^\xi, A^\xi, I^\xi, O^\xi, \gamma^\xi, \tau^\xi, \iota^\xi,
o^\xi,\mu_0^\xi,C^\xi,F^\xi,G^\xi).
\end{equation}

The rest of the section describes how its element are derived from the
\ac{SAN-T} specification.
We separate the presentation of the algorithm in two
parts:
\begin{inparaenum}[i)]
	\item concretization of the individual places, markings, and gates; and
	\item concretization of the overall model structure.
\end{inparaenum}
In all the following definitions, $\xi$ is the assignment of parameters from
which the instance should be generated.


\subsection{Places, Marking, and Gates}

\subsubsection{Places}

For each \emph{place template} in the \ac{SAN-T} model, one or more ``normal''
places are created in the instance model.
How many places are created, and with which indices, is given by applying the
assignment function on the multiplicity of the place template.

Formally, given a place template $\tilde
p=(\tau,k)\in \tilde{P}$ of the \ac{SAN-T} model,
and being $\mathit{Val}_\xi(k)=\{a_1,\ldots,a_{m}\}$ the indices
obtained from applying the assignment function to the multiplicty specification,
the places
$\{\tau_{a_1}^\xi,\ldots, \tau_{a_m}^\xi\}$ are created in
the concrete SAN model.

%
In the following we denote with
${\Pi}(\tilde p,i)\in P^\xi$ the $i$-th concrete place originating from
place template $\tilde p$. That is, ${\Pi}(\tilde p,i)=\tau_{a_i}^\xi$.

\subsubsection{Marking}

From any given \emph{marking} of a \ac{SAN-T} model a unique mapping to a
marking of the generated SAN instance can be identified.
Essentially, this is done by applying the marking template function to the index
of the generated places.

Formally, given a marking of the \ac{SAN-T} model, $\tilde{\mu}\in\tilde
M_{\tilde{P}}$, the marking $\mu^{\xi}\in M_P$ of the instance model is defined
as:
\begin{equation}
\mu^{\xi}(\tau_{a_i})=f_{\tilde{p}}(a_i),\qquad\forall \tilde
p\in\tilde P,\;\forall i\in\mathbb{N},
\end{equation}
where:
\begin{equation}
\tau_{a_i}=\Pi(\tilde p,i), \qquad f_{\tilde{p}}=\mu(\tilde{p}).
\end{equation}

That is, the marking of the $i$-th place ($\tau_{a_i}$) generated from place
template $\tilde p$  is obtained by applying \emph{marking template} function
($f_{\tilde{p}}$) to the index of the concrete place ($a_i$).

Given a marking $\tilde\mu\in\tilde{M}_{\tilde{P}}$ of the \ac{SAN-T} model, we
denote the corresponding marking $\mu^\xi\in M_P$ of the generated instance as
${\Gamma}(\tilde{\mu})$. Conversely, given a marking $\mu^{\xi}$ of the concrete
(generated) SAN model we denote as $\Gamma^{-1}(\mu^{\xi})$ the corresponding
marking $\tilde\mu$ of the originating \ac{SAN-T}.

\subsubsection{Input Gates}

Each \emph{input gate template} of the \ac{SAN-T} model is translated to exactly
one input gate in the \ac{SAN} instance. 
Given an input gate template $\tilde{g}=(\tilde{G},\tilde{e},\tilde{f})\in
\tilde{I}$,  we denote with
${\alpha}(\tilde{g})$ the corresponding input gate
$g^{\xi}=(G^{\xi},e^{\xi},f^{\xi})\in I^{\xi}$ in the concrete SAN model,
which is obtained as:
\begin{equation*}
\begin{split}
G^{\xi} &= \left\{ \Pi(\tilde p,j)\mid \tilde p=(\tau,k)\in
\tilde{G},\;j\in \mathit{Val}_\xi(k) \right\},
\\
e^{\xi}(\Gamma(\tilde\mu))&=\mathit{Val}_\xi(\tilde
e\left(\tilde{\mu})\right),\\
f^{\xi}(\Gamma(\tilde\mu))&=\tilde
f\left(\tilde{\mu},\xi\right). 
\end{split}\stepcounter{equation}\tag{\theequation}
\end{equation*}

That is, the input places of the concrete input gate, $G^\xi$, are all the
places generated from input place templates in $\tilde G$;
the input predicate applied to a marking $\Gamma(\tilde\mu)$ is the result of
applying the assignment function to the predicate of the gate template;
and the input function applied to marking $\Gamma(\tilde\mu)$ is the input
function of the gate template applied on marking $\tilde\mu$ and assignment
$\xi$.


\subsubsection{Output Gates}

Differently from input gate templates, each \emph{output gate template} may be
expanded to one or more concrete output gates.
The number of concrete output gates that should be generated depends on
parameters, and more specifically, from the parameter that controls the number
of cases of the connected activity.

 Given an output gate template
$(\tilde{G},\tilde{f})\in \tilde O$, 
we denote with ${\beta}(\tilde g,i)$ the $i$-th output gate
$(G_i^{\xi},f_i^{\xi})\in O$ generated from it in the SAN model, which
is obtained as:
\begin{equation*}
\label{eq:concretize-og}
\begin{split}
G_i^{\xi} &= \left\{ \Pi(\tilde p,j)\mid \tilde p=(\tau,k)\in
\tilde{G},\;j\in \mathit{Val}_\xi(k) \right\},
\\
f_i^{\xi}(\Gamma(\tilde\mu))&=\tilde
f\left(\tilde\mu,i,\xi\right).
\end{split}\stepcounter{equation}\tag{\theequation}
\end{equation*}

That is, the output places $G^\xi$ are all the places generated from
output place templates in $\tilde G$, 
and the output function applied to marking $\Gamma(\tilde\mu)$ is the
output function of the gate template applied on marking $\tilde\mu$, index $i$, and
assignment $\xi$.

\subsection{Overall SAN Definition}

We can now provide the complete specification of the SAN derived from a
\ac{SAN-T} $S_\Delta$ and an assignment $\xi$. That is, we can precisely
define all the elements in \autoref{eq:sant-derivation}, as follows:
%
%
\begin{equation}
\label{eq:concretize-overall}
\small
\begin{split}
P^\xi&=
	\bigcup_{\tilde p=(\tau,k)\in\tilde{P}}{
		\left\{\Pi(\tilde p,i)\mid i\in
		\mathit{Val}_\xi(k)\right\} };
\\
A^\xi&=\tilde{A};
\\
\gamma^\xi(a)&=\mathit{Val}_\xi(\tilde\gamma\left(\tilde a\right));
\\
I^\xi&=\left\{\alpha(\tilde g)\mid\tilde g\in \tilde{I}\right\};
\\
O^\xi&={\bigcup_{\tilde
g\in\tilde{O}}\left\{\beta(\tilde g,1),\ldots,\beta(\tilde
g,\mathit{Val}_\xi(\tilde\gamma\left(\tilde a\right)))\mid \tilde a=\tilde{o}(\tilde g)\right\}};
\\
\tau^\xi&=\tilde{\tau};
\\
\iota^\xi(\alpha(g))&=\tilde{\iota}(g),\quad\forall g\in\tilde{I};
\\
o^\xi(\beta(g,i))&=\tilde{o}(g),\quad\forall g\in \tilde{O},\forall
i\in
\left\{1,\ldots,\mathit{Val}_\xi(\tilde\gamma\left(\tilde a\right))\right\};
\\
\mu_0^\xi&=\tilde{\mu}_0(\xi).
\end{split}
\end{equation}

The rationale behind the above derivation can be summarized as follows:
\begin{inparaenum}[i)]
	\item the set of places $P^\xi$ is given by all the places derived from all the
	place templates in $\tilde P$;
	\item the set of activities remains unchanged;
	\item the function $\gamma$, which specifies the number of cases of an
	activity, is the result of applying the assignment function to the $\tilde\gamma$
	function;
	\item there is an input gate in $I^\xi$ for each input gate template in
	$\tilde I$;
	\item each output gate template in $\tilde O$ is expanded to a certain number of output
	gates, given by the number of cases of the activity to which it is connected; 
	\item the function $\tau$ that determines if an activity is timed or
	instantaneous remains unchanged;
	\item if an input gate template is connected to an activity template, then its
	concrete projection is connected to the projection of the activity template;
	\item if an output gate template is connected to an activity template,
	then all its concrete projections are connected to the projection of the
	activity template; and
	\item the initial marking $\mu_0^\xi$ is given by the initial marking of the
	\ac{SAN-T} model, applied to the assignment $\xi$.
\end{inparaenum}

Furthermore:
\begin{itemize}
  \item For each function $\tilde C_{\tilde a}$ in the case
  distribution assignment $\tilde{C}$, a corresponding function $C_a^\xi$ is
  included in $C^\xi$, defined as
  $C_a^\xi\left(\Gamma{(\mu)},k\right)=\tilde
  C_{\tilde a}\left(\mu,k,\xi\right),
  \;\forall\mu\in M_{\tilde{P}},\;\forall k\in\mathbb{N^+}$.
  \item For each function $\tilde F_{\tilde a}$ in the activity time
  distribution assignment $\tilde{F}$, a corresponding function $F_a^\xi$ is
  included in $F^\xi$, defined as
  $F_a^\xi(\Gamma{(\mu)},r)=\tilde F_{\tilde a}(\mu,r,\xi), \;\forall\mu\in
  M_{\tilde{P}},\;\forall r\in\mathbb{R}$.
  \item For each function $\tilde G_{\tilde a}$ in the
  reactivation function assignment $\tilde{G}$, a corresponding function
  $G_a^\xi$ is added to $G^\xi$, defined as $G_a^\xi(\Gamma{(\mu)})=\big\{\Gamma(\tilde\mu)\mid \tilde\mu\in\tilde
  G_a(\tilde\mu)\big\}\; \forall\mu\in\tilde M_{\tilde S}$.
\end{itemize}

\subsection{Example: Instances of the \emph{User} \ac{SAN-T}}

Following the \emph{concretize} algorithm described in the previous section, we
show here how it is possible to derive multiple instance of the \emph{User}
\ac{SAN} Template defined in \autoref{eq:santfull} and depicted in
\autoref{fig:motivation-sant}.


In particular, we show how the two concrete SAN models of 
\autoref{fig:motivation-san1} and \autoref{fig:motivation-san2} can be derived
by different assignments of parameters.
We define two different assignment
functions, $\xi_\mathtt{UserInternal}$ and $\xi_\mathtt{UserPress}$, which
will result in the generation of the two SAN instances:
\begin{equation}
\begin{split}
\xi_\mathtt{UserInternal} &= \{(s,\{1,6,7\}), (p,\{0.7,0.2,0.1\})\},\\
\xi_\mathtt{UserPress} &= \{(s,\{3,7\}), (p,\{0.6,0.4\})\}.
\end{split}
\end{equation}

The generation of the two instances follows a similar process, and for this
reason we show only one of them in details. The SAN model \emph{UserInternal}
is derived by the template-assignment pair
$(S_{\mathtt{User}},\xi_{\mathtt{UserInternal}})$, resulting in the following
\ac{SAN} model, where $\mathtt{ActivityName}(k)$ denotes the $k$-th case of the
(concrete) activity $\mathtt{ActivityName}$.

\begin{small}
\begin{align*}
\label{eq:san-instance-usernormal}
\text{SAN}_{\mathtt{UserInternal}} &=
(P^\xi, A^\xi, I^\xi, O^\xi, \gamma^\xi, \tau^\xi, \iota^\xi,
o^\xi,\mu_0^\xi,C^\xi,F^\xi,G^\xi)\\
{P^\xi}&=\{\mathtt{Idle}_1,\mathtt{Req}_1,\mathtt{Req}_6,\mathtt{Req}_7,\mathtt{Dropped}_1,\mathtt{Failed}_1\}\\
{A^\xi}&=\{\mathtt{Request},\mathtt{Fail},\mathtt{Drop}\}\\
{I^\xi}&=\{\mathtt{IGRequest},\mathtt{ArcInFail},\mathtt{ArcInDrop}\}\\
{O^\xi}&=\{\mathtt{OGRequest}_1,\mathtt{OGRequest}_2,\mathtt{OGRequest}_3,\\
&\phantom{=\ \{}
\mathtt{ArcOutFail},\mathtt{ArcOutDrop}\}\\
{\gamma^\xi} &= \{ (\mathtt{Request},3), (\mathtt{Fail},1),
(\mathtt{Drop},1)\}\\
{\tau^\xi} &= \{ (\mathtt{Request},\mathrm{\emph{timed}}),
(\mathtt{Fail},\mathrm{\emph{instantaneous}}),\\
&\phantom{=\ \{}
(\mathtt{Drop},\mathrm{\emph{instantaneous}})\}\\
{\iota^\xi}&=\{(\mathtt{IGRequest},\mathtt{Request}),
(\mathtt{ArcInFail},\mathtt{Fail}),\\
&\phantom{=\ \{}(\mathtt{ArcInDrop},\mathtt{Drop})\}\\
{o^\xi}&=\{(\mathtt{OGRequest}_1,\mathtt{Request}(1)),\\
&\phantom{=\ \{}(\mathtt{OGRequest}_2,\mathtt{Request}(2)),\\
&\phantom{=\ \{}(\mathtt{OGRequest}_3,\mathtt{Request}(3)),\\
&\phantom{=\ \{}(\mathtt{ArcOutFail},\mathtt{Fail}(1)),\\
&\phantom{=\ \{}(\mathtt{ArcOutDrop},\mathtt{Drop}(1))\}\\
{\mu_0^\xi(p)}&=\left\{
	\begin{array}{ll}
	1 & \textrm{if }{p}=\mathtt{Idle}_1,\\
	0 & \textrm{otherwise}.
	\end{array}
\right.\\
{C^\xi}&=\{{C}_{\mathtt{Request}},{C}_{\mathtt{Drop}},{C}_{\mathtt{Fail}}\}\stepcounter{equation}\tag{\theequation}\\
{C}_{\mathtt{Request}}(\tilde{\mu},i)&=
\left\{
	\begin{array}{cl}
	0.7 & \text{if } i=1,\\
	0.2 & \text{if } i=2,\\
	0.1 & \text{if } i=3,\\
	0 & \mathrm{otherwise}.
	\end{array}
\right.\\
{C}_{\mathtt{Drop}}(\tilde{\mu},i)&=
{C}_{\mathtt{Fail}}(\tilde{\mu},i)=
\left\{
	\begin{array}{ll}
	1 & \text{if } i=1,\\
	0 & \mathrm{otherwise}.
	\end{array}
\right.\\
{F^\xi}&=\{{F}_{\mathtt{Request}}\},\qquad {G^\xi}=\{{G}_{\mathtt{Request}},{G}_{\mathtt{Drop}},{G}_{\mathtt{Fail}}\}.\\
\end{align*}
\end{small}
\vspace{-6mm}

Elements ${F^\xi}$ and ${G^\xi}$ are not discussed in details, since their
derivation is straightforward for this model. Also, as discussed before, the are
no reactivation markings and therefore the functions ${G}_{\mathtt{Request}}$,
${G}_{\mathtt{Drop}}$, and ${G}_{\mathtt{Fail}}$ are in this
case the empty function.

It is instead worth detailing the processing of the output gate template
\texttt{OGRequest}, which is affected by variability. In fact, in the \ac{SAN-T}
model it is connected to an activity template with a variable number of cases.

According to \autoref{eq:concretize-overall}, the number of concrete output
gates that are generated from each output gate template is given by
$\mathit{Val}_\xi(\tilde\gamma\left(\tilde a\right))$, where $\tilde{a}$ is the
associated activity template. In our case, 
$\mathit{Val}_{\xi_{\mathtt{UserInternal}}}(\tilde\gamma\left(\mathtt{Request}\right))=3$,
and therefore three output gates are created: \texttt{OGRequest}$_1$,
\texttt{OGRequest}$_2$, and \texttt{OGRequest}$_3$.
Each of these output gates is connected to the corresponding case of the
\texttt{Request} activity (see $o^\xi$ in
\autoref{eq:san-instance-usernormal}).

The definition of each output gate, in terms of its input places and output
function, is obtained by \autoref{eq:concretize-og}. For the \emph{UserInternal}
instance they are defined as follows.

\begin{small}
\begin{align*}
\mathtt{OGRequest}_1&= (G_1,f_1)\stepcounter{equation}\tag{\theequation},\qquad
G_1=\{\mathtt{Req}_1,\mathtt{Req}_6,\mathtt{Req}_7\}, \\
f_1(\mu)&=\mu'_1\;\forall \mu\in M_G\mid\mu'_1(p)=\left\{
\begin{array}{ll}
1 & \text{if }p=\mathtt{Req}_1, \\
\mu(p) & \text{otherwise}. \\
\end{array}
\right.\\
\mathtt{OGRequest}_2&= (G_2,f_2),\qquad G_2=\{\mathtt{Req}_1,\mathtt{Req}_6,\mathtt{Req}_7\}, \\
f_2(\mu)&=\mu'_2\;\forall \mu\in M_G\mid\mu'_2(p)=\left\{
\begin{array}{ll}
1 & \text{if }p=\mathtt{Req}_6, \\
\mu(p) & \text{otherwise}. \\
\end{array}
\right.\\
\mathtt{OGRequest}_3&= (G_3,f_3),\qquad G_3=\{\mathtt{Req}_1,\mathtt{Req}_6,\mathtt{Req}_7\}, \\
f_3(\mu)&=\mu'_3\;\forall \mu\in M_G\mid\mu'_3(p)=\left\{
\begin{array}{ll}
1 & \text{if }p=\mathtt{Req}_7, \\
\mu(p) & \text{otherwise}. \\
\end{array}
\right.\\
\end{align*}
\end{small}
Each of the three gates adds a token to the corresponding place generated from
place template \texttt{Req}, basically corresponding to the output
arcs appearing in \autoref{fig:motivation-san1}.

\section{Application to the \acs{BEDCS} Network }
\label{sec:casestudy}

In this section we show how the proposed \ac{SAN-T} formalization can be used to
model a real system.
We apply the formalism to the modeling of the 
\ac{BEDCS}, and in particular we discuss here the modeling of its backbone
network.
%

To demonstrate the generality of our formalism, we base the \ac{BEDCS}
network model on the work in \cite{Nencioni17}, in which the objective of the
authors was to evaluate a backbone network in Norway, detailing failure
correlation between system components.
The model in \cite{Nencioni17} is based on \acp{SAN}, and it has been defined in
a modular way as typically done in the literature.
However, as discussed in the motivations for the present work, in \cite{Nencioni17}
elements are only described by examples, because they include
variability aspects that cannot be represented using \acp{SAN} alone.

Here we show how the \ac{SAN-T} formalism can accurately
define such \ac{SAN}-based models, including their variability aspects. It
should be noted that the authors of \cite{Nencioni17} defined their models
without any collaboration with the authors of this paper, and they are not
involved in this work.

%

\subsection{Models of the Backend Network}

We focus here on the modeling of the ground sector, and in particular
of the backbone network connecting the ground stations.
In fact, the ground segment is currently composed of four ground stations
located in different states of Brazil:
two receiving stations, in Cuiabá (MT) and Alcântara (MA), the
mission center in Natal (RN), and the remote control
center in São José dos Campos (SP) \cite{deCarvalho13}. The two closest stations
are more than 1000 km apart.


As mentioned above, we adopt the approach of \cite{Nencioni17} to model the
\ac{BEDCS} backbone network. That work fits particularly well to our problem,
because the authors provided models for different architectural options,
including traditional network infrastructures, as well as those based on
\acp{SDN}. Furthermore, the model takes into account for different kinds of
correlation between component failures, due to e.g., physical proximity.

The modeling approach in \cite{Nencioni17} defines different kinds
of building blocks for the system model, grouped in two categories:
\emph{Component Blocks} and \emph{Dependency Blocks}.

The \emph{Component Blocks} are simple \ac{SAN} models that represent physical
components of the system architecture. In particular, the authors consider 
a ``template'' model for a \texttt{Link}, an \texttt{IPRouter}, a
\texttt{SDNSwitch}, and a \texttt{SDNController}. The variability in
these models is given only by the rates associated with the firing of
activities, and the probabilities associated with their cases.
Therefore, these models can be represented by plain \acp{SAN}.

The \emph{Dependency Blocks} are used to model the occurrence of dependent
failures between components. The model considers 7 kinds of dependencies
between components:
\begin{inparaenum}[i)]
\item {\em Geographical Proximity (GEO)}, when a small geographical distance 
results in common sensitivity to bad weather and
natural disasters;
\item {\em Physical Proximity (PHY)}, which causes a strong failure
correlation (e.g., blackout);
\item {\em Common O\&M (COM)}, in which the \ac{O+M} is
actually the same for multiple network elements;
\item {\em Misconfiguration (MIS)}, when elements
share the same configuration or have a correlated logic;
\item {\em Compatibility Issue (CIS)}, when a simultaneous failure may occur on
multiple network elements due to incompatibility issues among them;
\item {\em Homogeneous Equipment (HEQ)}, that is, when a failure happens in a
network element, another element with the same equipment may likely fail as well; and
\item {\em Traffic Migration (TMI)}, that is, when a network element fails, it
could happen that the replacement network element is not able to take over.
\end{inparaenum}

Specific \ac{SAN} models that represent these dependencies are ``plugged'' in
the overall system model, according to the scenario to be represented. While
these models have been described with examples in \cite{Nencioni17}, they have
variability that cannot be expressed with plain \acp{SAN}, for example in terms
of how many components, and which ones, are involved in the dependency.


\subsection{Modeling with \acp{SAN-T}}

We discuss here how \acp{SAN-T} can be used to formally specify such dependency
blocks, in particular for the GEO and TMI dependencies.
Models for the other dependencies can be defined in a similar way.

\subsubsection{GEO Dependency Block}

\begin{figure}
	\centering
	\includegraphics[scale=\sanscale]{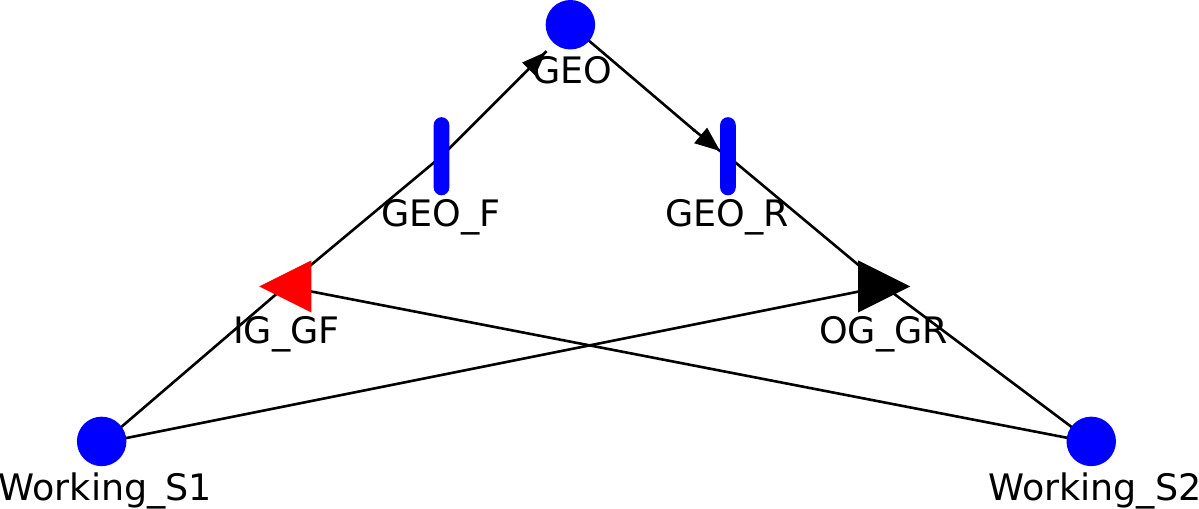}
	\caption{Example of the GEO building block specified with \acp{SAN}, for two
	components. Figure reproduced from \cite{Nencioni17}.}
	\label{fig:geo-san}
\end{figure}

The original GEO dependency block as defined in \cite{Nencioni17} is depicted in
\autoref{fig:geo-san}. The general idea of the block is as follows.

Places \texttt{Working\_S1} and \texttt{Working\_S2} represent the working state
of the two components involved in the dependency, in this case two \ac{SDN}
switches (S1 and S2). If place \texttt{Working\_SX} contains a token it means
that the corresponding component is currently working. When both components
are working, the activity \texttt{GEO\_F} is enabled, meaning that the GEO
common cause failure may occur.
Once the failure has occurred, restoration is possible after some time,
represented by the timed activity \texttt{GEO\_R}. Restoration makes the
involved components working again, by adding a token to the \texttt{Working\_SX}
places.

\begin{figure}
	\begin{subfigure}{.48\linewidth}
	\centering
	\includegraphics[scale=\sanscale]{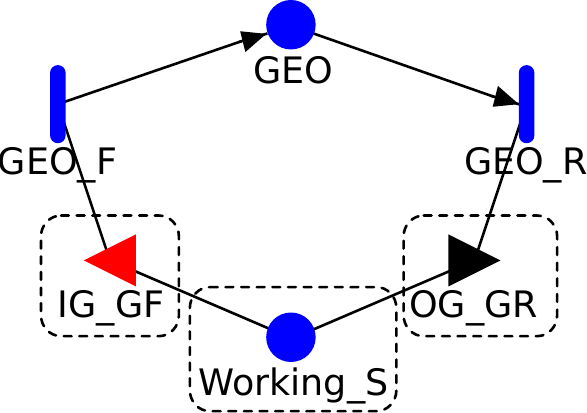}
	\caption{}
	\label{fig:geo-sant-normal}
	\end{subfigure}
	\hfill
	\begin{subfigure}{.48\linewidth}
	\centering
	\includegraphics[scale=\sanscale]{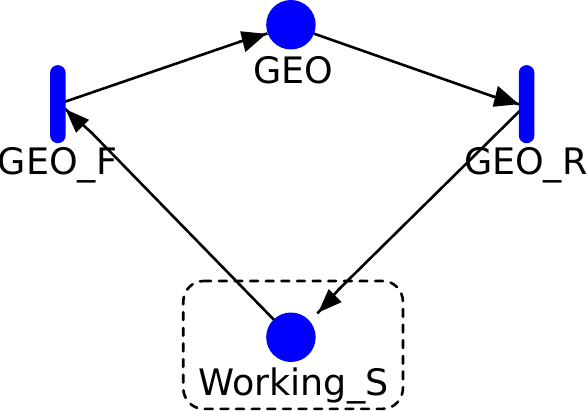}
	\caption{}
	\label{fig:geo-sant-witharcs}
	\end{subfigure}
\caption{\subref{fig:geo-sant-normal} Generalized GEO building block specified
with \ac{SAN-T}, for any number of components.
\subref{fig:geo-sant-witharcs} Simplified notation using input arc templates and
output arc templates.}
\label{fig:geo-sant}
\end{figure}

This block has been defined, as an example, for two components only.
However, the GEO dependency may involve three or more switches and, in general,
any number of components. The block can be generalized, informally as follows:
\emph{``For each switch $X$ involved in the dependency create a place
\texttt{Working\_SX}. The enabling predicate of the input gate
\texttt{IG\_GF} is true when all the places \texttt{Working\_SX} contain a
token, and the input function removes all the tokens from those
places. The output function of gate \texttt{OG\_GR} adds a token to all the
\texttt{Working\_SX} places.''}

Using the proposed \ac{SAN-T} formalism, the generalized ``template'' version of
the block can be defined in a precise way, as follows. The corresponding
graphical representation of the model is depicted in
\autoref{fig:geo-sant-normal} using gates, and in
\autoref{fig:geo-sant-witharcs} using the compact notation with arc templates.
The \ac{SAN-T} model has three parameters: $n$, the identifiers of the
components involved in the dependency; $\lambda^f$, the rate of occurrence of
the GEO failure; and $\lambda^r$, the restoration rate.

\begin{small}
\begin{align*}
\label{eq:geo}
\text{SAN-T}_{\mathtt{GEO}} &= (\Delta,\tilde{P},\tilde{A},\tilde{I},\tilde{O},
\tilde{\gamma}, \tilde{\tau}, \tilde{\iota},\tilde{o}, \tilde\mu_0,\tilde
C,\tilde F,\tilde G)\\
\Delta&=\{n_{\tintset},{\lambda^f}_{\treal},{\lambda^r}_{\treal}\}\\
\tilde{P}&=\{(\mathtt{GEO},1),(\mathtt{Working\_S},n)\}\\
\tilde{A}&=\{\mathtt{GEO\_F},\mathtt{GEO\_R}\}\\
\tilde{I}&=\{\mathtt{IG\_GF},\mathtt{GEOtoGEO\_R}\}\\
\tilde{O}&=\{\mathtt{OG\_GR},\mathtt{GEO\_FtoGEO}\}\\
\tilde{\gamma} &= \{ (\mathtt{GEO\_F},1), (\mathtt{GEO\_R},1)\}\\
\tilde{\tau} &= \{ (\mathtt{GEO\_F},\mathrm{\emph{timed}}),
(\mathtt{GEO\_R},\mathrm{\emph{timed}})\}\\
\tilde{\iota}&=\{(\mathtt{IG\_GF},\mathtt{GEO\_F}),(\mathtt{GEOtoGEO\_R},\mathtt{GEO\_R})\}\\
\tilde{o}&=\{(\mathtt{OG\_GR},\mathtt{GEO\_R}),(\mathtt{GEO\_FtoGEO},\mathtt{GEO\_F})\}\\
\tilde{\mu}_0(\xi)&=\mu'(\tilde{p})\;\forall\xi\mid\mu'(\tilde{p})=\left\{
	\begin{array}{ll}
	f^1 & \textrm{if }\tilde{p}=\mathtt{Working\_S},\\
	f^0 & \textrm{otherwise}.
	\end{array}
\right.\\
\tilde{C}&=\{\tilde{C}_{\mathtt{GEO\_F}},\tilde{C}_{\mathtt{GEO\_R}}\}\stepcounter{equation}\tag{\theequation}\\
\tilde{C}_{\mathtt{GEO\_F}}(\tilde{\mu},i,\xi)&=
\tilde{C}_{\mathtt{GEO\_R}}(\tilde{\mu},i,\xi)=
\left\{
	\begin{array}{ll}
	1 & \text{if } i=1,\\
	0 & \mathrm{otherwise}.
	\end{array}
\right.\\
\tilde{F}&=\{\tilde{F}_{\mathtt{GEO\_F}},\tilde{F}_{\mathtt{GEO\_R}}\},\qquad
\tilde{G}=\{\tilde{G}_{\mathtt{GEO\_F}},\tilde{G}_{\mathtt{GEO\_R}}\}.\\
\end{align*}
\end{small}
\vspace{-5mm}

As in the previous examples, the $\tilde{G}_{\mathtt{GEO\_F}}$ and
$\tilde{G}_{\mathtt{GEO\_F}}$ functions are empty, as none of the activities is
reactivating. The firing distributions are negative exponential distributions
based on the $\lambda^f$ and $\lambda^r$ parameters, that is,
$\tilde{F}_{\mathtt{GEO\_F}}(t)=1-e^{-\lambda^{f}\cdot t}$ and
$\tilde{F}_{\mathtt{GEO\_R}}(t)=1-e^{-\lambda^{r}\cdot t}$.
The gates \texttt{GEO\_FtoGEO} and \texttt{GEOtoGEO\_R} correspond to the
two arcs incoming to and outgoing from the \texttt{GEO} place, respectively, which 
are ``normal'' arcs as in ordinary \acp{SAN}. Conversely, the gates \texttt{IG\_GF} and
\texttt{OG\_GR} are where the variability in the behavior resides, and
its definition is thus detailed in the following.

\begin{small}
\begin{align*}
\mathtt{IG\_GF}&=(\tilde{G}_{\mathtt{IF\_GF}},\tilde{e}_{\mathtt{IF\_GF}},\tilde{f}_{\mathtt{IF\_GF}}),\\
\tilde{G}_{\mathtt{IG\_GF}}&=\{\mathtt{Working\_S}\},\\
\tilde{e}_{\mathtt{IG\_GF}}(\tilde{\mu},\xi)&=\bigwedge_{i\in\mathit{Val}_\xi(n)}\Bigl(\tilde{\mu}_{\tilde{p}}(i)>0\Bigr),\quad\text{with
}\tilde{\mu}_{\tilde{p}}=\tilde\mu(\tilde p),\\
\tilde{f}_{\mathtt{IG\_GF}}(\tilde{\mu},i,\xi)&=\mu'(\tilde{p})\;\forall\tilde
p\in\tilde P\mid\\
&\phantom{=\ \ }\mu'(\tilde{p})= \left\{
	\begin{array}{ll}
	f^0  & \textrm{if
	}\tilde{p}=\mathtt{Working\_S},\\
	\tilde{\mu}(\tilde{p}) & \textrm{otherwise}.
	\end{array}
\right.\\
\\\stepcounter{equation}\tag{\theequation}
\mathtt{OG\_GR}&=(\tilde{G}_{\mathtt{OG\_GR}},\tilde{f}_{\mathtt{OG\_GR}}),\\
\tilde{G}_{\mathtt{OG\_GR}}&=\{\mathtt{Working\_S}\},\\
\tilde{f}_{\mathtt{OG\_GR}}(\tilde{\mu},i,\xi)&=\mu'(\tilde{p})\;\forall\tilde
p\in\tilde P\mid\\
&\phantom{=\ \ }\mu'(\tilde{p})= \left\{
	\begin{array}{ll}
	f^1  & \textrm{if
	}\tilde{p}=\mathtt{Working\_S},\\
	\tilde{\mu}(\tilde{p}) & \textrm{otherwise}.
	\end{array}
\right.\\
\end{align*}
\end{small}

\subsubsection{TMI Dependency Block}

\begin{figure}
	\centering
	\includegraphics[width=\linewidth]{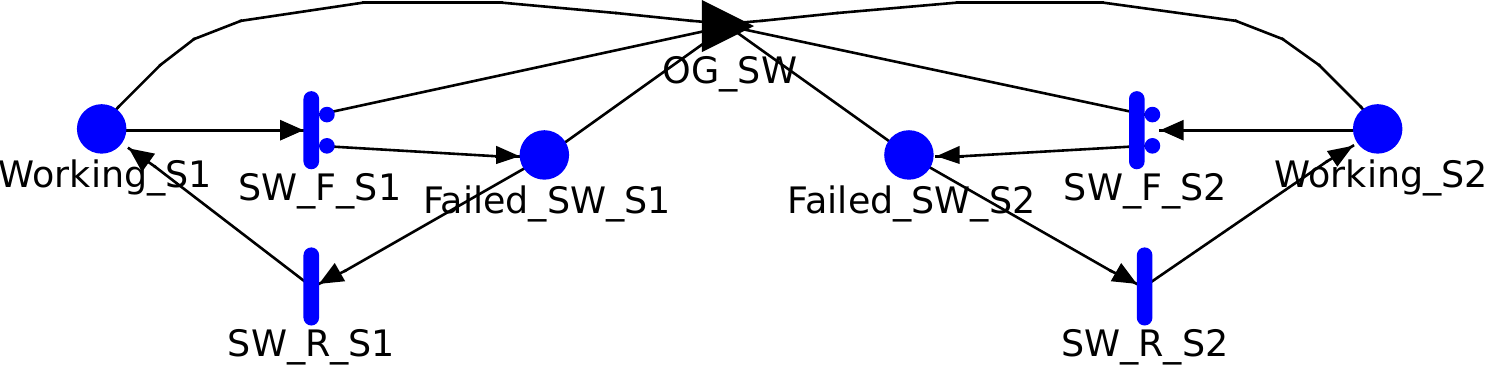}
	\caption{Example of the TMI dependency between two switches S1 and S2, modeled
	with \acp{SAN}. Figure reproduced from \cite{Nencioni17}.}
	\label{fig:tmi}
	\vspace{-3mm}
\end{figure}

The example in \autoref{fig:tmi} shows the introduction of the TMI dependency
among two \ac{SDN} switches. The idea is that upon software failure of one of
them, there is a probability that traffic migration also causes
the second switch to fail.
This dependency does not actually add a new block to the system model, but
instead it modifies the existing \ac{SAN} models of the involved components.

The model in \autoref{fig:tmi} shows the \ac{SAN} models of the
two switches (S1 on the left, S2 on the right), and a new output gate
\texttt{OG\_SW} that represents the dependency.
With respect to the normal model of the switch, a new case is added to the
failure activity (\texttt{SW\_F\_S1} and \texttt{SW\_F\_S2}), and it is
connected to the newly introduced \texttt{OG\_SW} output gate.
It should be noted that the figure contains a slight abuse of notation (as in
the original paper), because in principle an output gate can be connected to
only one activity. We consider therefore two identical
copies of the gate, each connected to one of the two activities.

\begin{figure}
	\centering
	\includegraphics[scale=\sanscale]{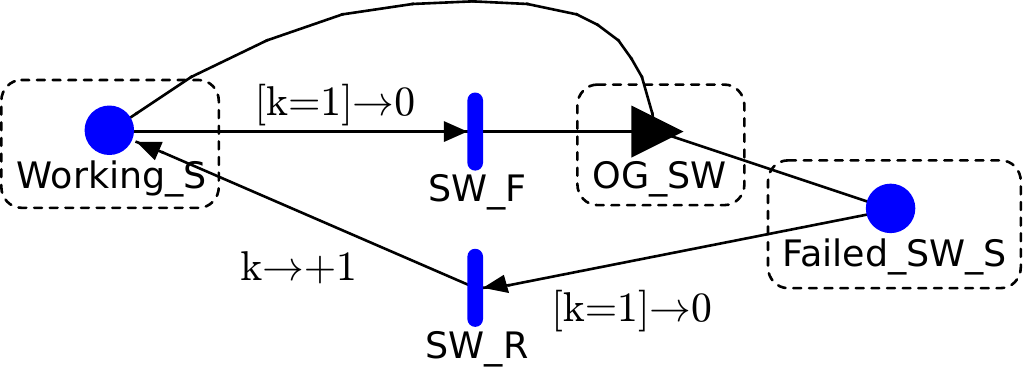}
	\caption{SAN-T model of the \ac{SDN} switches considering the TMI dependency
	in a general way.}
	\label{fig:tmi-sant}
	\vspace{-3mm}
\end{figure}
As in the previous case, this dependency can span multiple switches; more in
general, each switch can affect a different subset of the switches in the
system. This dependency can be generalized and formalized by modeling the
\ac{SDN} switch as a \ac{SAN-T}. Which switches will be affected by the TMI dependency, and
whether the dependency must be represented at all, will be specified by
the parameters of the template. A graphical representation of the resulting
\ac{SAN-T} model is provided in \autoref{fig:tmi-sant}, while its formal
specification is provided in the following. 

\begin{small}
\begin{align*}
\label{eq:tmi}
\text{SAN-T}_{\mathtt{SwitchTMI}} &=
(\Delta,\tilde{P},\tilde{A},\tilde{I},\tilde{O}, \tilde{\gamma}, \tilde{\tau}, \tilde{\iota},\tilde{o}, \tilde\mu_0,\tilde
C,\tilde F,\tilde G)\\
\Delta&=\{k_\tint,J_{\tintset},p^{\mathtt{TMI}}_\treal,{\lambda^f}_{\treal},{\lambda^r}_{\treal}\}\\
\tilde{P}&=\{(\mathtt{Working\_S},J\cup\{k\}),
(\mathtt{Failed\_SW\_S},J\cup\{k\})\}\\
\tilde{A}&=\{\mathtt{SW\_F},\mathtt{SW\_R}\}\\
\tilde{I}&=\{\mathtt{Working\_StoSW\_F},\mathtt{Failed\_SW\_StoSW\_R}\}\\
\tilde{O}&=\{\mathtt{OG\_SW},\mathtt{SW\_RtoWorking\_S}\}\\
\tilde{\gamma} &= \{ (\mathtt{SW\_F},1+(p^{\mathtt{TMI}}_\treal>0)),
(\mathtt{SW\_R},1)\}\\
\tilde{\tau} &= \{ (\mathtt{SW\_F},\mathrm{\emph{timed}}),
(\mathtt{SW\_R},\mathrm{\emph{timed}})\}\\
\tilde{\iota}&=\{(\mathtt{Working\_StoSW\_F},\mathtt{SW\_F}),\\
&\phantom{=\ \{}(\mathtt{Failed\_SW\_StoSW\_R},\mathtt{SW\_R})\}\\
\tilde{o}&=\{(\mathtt{OG\_SW},\mathtt{SW\_F}),(\mathtt{SW\_RtoWorking\_S},\mathtt{SW\_R})\}\\
\tilde{\mu}_0(\xi)&=\mu'(\tilde{p})\;\forall\xi\mid\mu'(\tilde{p})=\left\{
	\begin{array}{ll}
	f^1 & \textrm{if }\tilde{p}=\mathtt{Working\_S},\\
	f^0 & \textrm{otherwise}.
	\end{array}
\right.\\
\tilde{C}&=\{\tilde{C}_{\mathtt{SW\_F}},\tilde{C}_{\mathtt{SW\_R}}\}\stepcounter{equation}\tag{\theequation}\\
\tilde{C}_{\mathtt{SW\_F}}(\tilde{\mu},i,\xi)&=
\left\{
	\begin{array}{ll}
	1-p^{\mathtt{TMI}}_\treal & \text{if } i=1,\\
	p^{\mathtt{TMI}}_\treal & \text{if } i=2,\\
	0 & \mathrm{otherwise}.
	\end{array}
\right.\\
\tilde{C}_{\mathtt{SW\_R}}(\tilde{\mu},i,\xi)&=
\left\{
	\begin{array}{ll}
	1 & \text{if } i=1,\\
	0 & \mathrm{otherwise}.
	\end{array}
\right.\\
\tilde{F}&=\{\tilde{F}_{\mathtt{SW\_F}},\tilde{F}_{\mathtt{SW\_R}}\},\qquad
\tilde{G}=\{\tilde{G}_{\mathtt{SW\_F}},\tilde{G}_{\mathtt{SW\_R}}\}.\\
\end{align*}
\end{small}
\vspace{-5mm}

The template has five parameters: $k$, the index of the switch represented by
the instance; $J$, an array of identifiers of other switches that can be
affected when the switch fails; $p^{\mathtt{TMI}}$, the
probability that the TMI dependency occurs; and $\lambda^f$ and $\lambda^r$ as
failure and repair rates of the switch, respectively.

Places \texttt{Working\_S} and \texttt{Failed\_SW\_S} are template places, and
their multiplicity is given by the union of index of the switch and those of the
switches that should be affected by the TMI failure. That is, the generated
\ac{SAN} instances would contain a place \texttt{Working\_S}$k$ and a place
\texttt{Failed\_SW\_S}$k$ for the switch represented by the instance, and a 
place \texttt{Working\_S}$j$ and \texttt{Failed\_SW\_S}$j$ for each other switch
$j\in J$ that can be affected by the dependency.

Gates \texttt{Working\_StoSW\_F}, \texttt{Failed\_SW\_StoSW\_R}, and
\texttt{SW\_RtoWorking\_S} correspond to the arc templates depicted in the
figure. Their specification is given using the notation introduced in 
\autoref{sec:arc-templates} and thus they are not further
detailed here.
The specification of output gate template \texttt{OG\_SW} is detailed in the
following. Basically, for the first case of the activity it adds one token only
to the place with the same index as parameter $k$, and for the second case it
also adds a token to the places corresponding to the affected switches, and
removes the token from their $\texttt{Working\_S}$ place.
\begin{equation}
\small
\begin{split}
\mathtt{OG\_SW}&=(\tilde{G}_{\mathtt{OG\_SW}},\tilde{f}_{\mathtt{OG\_SW}}),\\
\tilde{G}_{\mathtt{OG\_SW}}&=\{\mathtt{Working\_S},\mathtt{failed\_SW\_S}
\},\\
\tilde{f}_{\mathtt{OG\_SW}}(\tilde{\mu},i,\xi)&=\mu'_i(\tilde{p})\;\forall\tilde
p\in\tilde P\mid\\
\mu'_1(\tilde{p})&= \left\{
	\begin{array}{ll}
	f_k^1  & \textrm{if }\tilde{p}=\mathtt{Failed\_SW\_S},\\
	\tilde{\mu}(\tilde{p}) & \textrm{otherwise}.
	\end{array}
\right.\\
\mu'_2(\tilde{p})&= \left\{
	\begin{array}{ll}
	f^0_{n}(\tilde p)  & \textrm{if }\tilde{p}=\mathtt{Working\_S},\\
	f^1_{\{k\}\cup J}(\tilde p)  & \textrm{if }\tilde{p}=\mathtt{Failed\_SW\_S},\\
	\tilde{\mu}(\tilde{p}) & \textrm{otherwise}.
	\end{array}
\right.\\
\end{split}
\end{equation}

\section{The TDML Framework}
\label{sec:tmdlframework}

The work in this paper complements the 
\ac{TMDL} framework that we defined in \cite{2020TR}. In this section we briefly
recall it, and discuss the relation with the work in this paper.

\begin{figure}
\centering
\includegraphics[width=\linewidth]{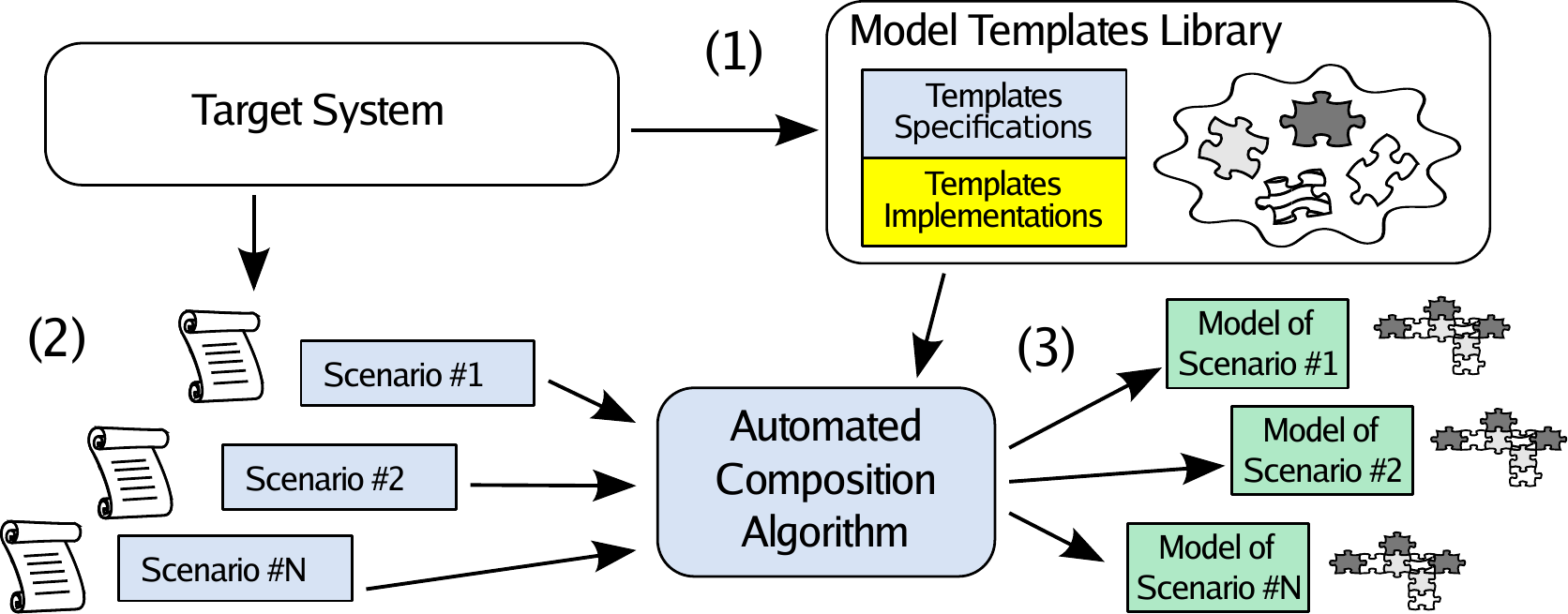}
\caption{Workflow of the TMDL framework for the automated generation of
performability models \cite{2020SERENE}.}
\label{fig:tmdl-workflow}
\vspace{-4mm}
\end{figure}

The idea behind the \ac{TMDL} framework is
organized in three steps:
\begin{inparaenum}[i)]
\item there exist a \emph{library} of parametric reusable submodels,
defined with a {template-level formalism}, and called \emph{model
templates};
\item based on the scenario to be modeled, a set of templates is
selected and proper parameters are assigned; and
\item models in the instance-level formalism are automatically
generated and assembled to obtain the overall system model.
\end{inparaenum}
The corresponding workflow is detailed in \autoref{fig:tmdl-workflow}.

In Step~\#1, a \emph{library} of reusable \emph{model templates} is created by
an expert.
In Step~\#2, the different system configurations that should be analyzed are
defined in terms of ``scenarios''. Scenarios are composed of \emph{model
variants}, that is, a selection of model templates with their parameter
assignment. 
In Step~\#3 all the needed \emph{model instances} are automatically created and
assembled, thus generating the complete system model for each scenario.
Note that the steps in the workflow are not
strictly sequential. In particular, the creation of the model
library is performed once, and the library is stored for future access.

What makes the model templates reusable is that they have well-defined
\emph{interfaces} and \emph{parameters}. Briefly summarizing, interfaces specify
how they can be connected to other templates, while parameters make it possible
to derive different concrete models from the same template.
A model template has a \emph{specification} (of its parameters and interfaces),
and an \emph{implementation}. 

The specification of a template is provided with the TMDL, a
\ac{DSL} specifically defined for that purpose.
The implementation of a template can be \emph{atomic} or \emph{composite}. A
composite implementation simply specifies which other templates can be composed
and how, and it is also specified with the \ac{TMDL}.
The implementation of an atomic template should be given using a \emph{template-level formalism},
that is, a modeling formalism that defines partially specified models. 
Conversely, we call
\emph{instance-level formalism} the modeling formalism concretely used for the
analysis, generated in
{Step~\#3} (e.g., ``normal'' \acp{SAN}).




In \cite{2020TR} we introduced some assumptions, both for
simplicity but also to keep the approach independent of a specific modeling formalism. In particular,
we assumed that for a certain instance-level formalism (e.g., \acp{SAN}) it was possible to
define:
\begin{inparaenum}[i)]
  \item a corresponding \emph{template-level formalism}, to specify model
  templates;
  \item a \emph{concretize} function that, given a model in the
  template-level formalism and an assignment of values to its parameters, generates a
  model in the instance-level formalism; and
\item a notion of \emph{compatibility} between the \ac{TMDL} specification of
  a template (i.e., interfaces and parameters) and its implementation with the
  template-level formalism.
\end{inparaenum}

In this paper we have provided a formal definition of \aclp{SAN-T}
(\acp{SAN-T}), and of the corresponding concretize function.
These definitions enable the
application of the \ac{TMDL} framework considering \acp{SAN} as the instance-level formalism.
%
%

\section{Conclusions}

\label{sec:conclusions}
In this paper we proposed a formal definition of Stochastic Activity Networks
Templates (SAN-T), a formalism that generalizes \acp{SAN} with the addition
of variability aspects.
\ac{SAN-T} models define abstract models depending on parameters, from which
concrete \ac{SAN} models can be generated by assignment of values.
Differently from other approaches, in \acp{SAN-T} parameters may also affect
the \emph{structure} of the model, e.g., the number of places in the model or the number of
cases of an activity. This formalism aims to simplify the specification of
generic building blocks for complex models based on \ac{SAN}, which is a common
problem in the literature.

We demonstrated the applicability of the formalism by using it for the
generalization of \ac{SAN}-based models present in the literature.
The proposed formalization can accurately describe the variation points present
in the models, and at the same time provide a compact notation, thanks to the
proposed extensions to the graphical notation of \acp{SAN}.

This work complements our work in \cite{2020TR}, in which we defined an
approach to simplify the composition of models based on \acp{SPN}.
As current and future work, we are working on two parallel directions. The first
is to apply the methodology for the actual evaluation of a real system.
In this paper we showed parts of the model of the \ac{BEDCS} systems, a
large-scale infrastructure for environmental monitoring in Brazil.
Further work is ongoing, in collaboration with the National Institute of Space
Research of Brazil (INPE), within the ADVANCE project \cite{ADVANCE}.

We are also working on providing tool support for the proposed formalism, to
facilitate its application by other researchers.
A prototype editor for \acp{SAN-T} models, based on the \ac{EMF}
and the Sirius
modeling tools is being developed as open source
software. The architecture of the editor and tool under development has been
presented in \cite{2020SERENE}.

\bibliographystyle{IEEEtran}
\bibliography{sant-paper}

\end{document}